\documentclass[fleqn,usenatbib]{mnras}

\usepackage[T1]{fontenc}

\DeclareRobustCommand{\VAN}[3]{#2}
\let\VANthebibliography\thebibliography
\def\thebibliography{\DeclareRobustCommand{\VAN}[3]{##3}\VANthebibliography}

\usepackage{graphicx}	
\usepackage{amsmath}	
\usepackage{amssymb}	
\usepackage{siunitx}
\usepackage{subcaption}
\usepackage{soul}
\usepackage{ulem}

\title{Variability of the Faraday Rotation Measure in the Parsec-scale Jet of AGN 1633+382}

\author[K. C. Lai et al.]{
K. C. Lai,$^{1}$
J. C. Algaba,$^{1}$\thanks{E-mail: algaba@um.edu.my}
Z. Z. Abidin$^{1}$
\\
$^{1}$Department of Physics, Universiti Malaya, 50603 Kuala Lumpur, Malaysia\\
}

\date{Accepted XXX. Received YYY; in original form ZZZ}

\pubyear{2024}

\begin{document}
\label{firstpage}
\pagerange{\pageref{firstpage}--\pageref{lastpage}}
\maketitle

\begin{abstract}
We study the Faraday rotation measure (RM) variability of flat spectrum radio quasar 1633+382 on 5 epochs spanning from 2004 to 2008. We used 4 to 43~GHz VLBI polarization data from VLBA. Core RM across 4 to 15 GHz scales with a power index $a\sim2$, for $a$ in RM$\propto \nu^a$. We detected sign changes across epochs, both in the core and in the jet region. RM time variability in the core and jet region are not correlated, hence limiting the size of a possible Faraday screen. We relate the core RM variability to a new component emerging from the core region. For the jet region, we consider the jet-medium interaction to be a less likely cause of the RM variability because of the uniform spectral index distribution. The observed RM value variation requires a huge fluctuation in electron density or magnetic field, hence a foreground Faraday screen is less favoured. We further discuss other possibilities of the RM variability based on jet kinematics. 
\end{abstract}




\section{Introduction}

Polarized emission from active galactic nuclei (AGNs) jets provides crucial information about the jet environment. When linearly polarized emission passes through a thermal magnetized plasma, the polarization angle is rotated. This effect is known as Faraday rotation. The amount of rotation is proportional to the square of wavelength ($\lambda^2$), so the observed electric vector position angle (EVPA) is given by
\begin{equation}
	\chi_{\text{obs}} = \chi_{\text{int}} + \text{RM} \times \lambda^2
\end{equation}
where $\chi_{\mathrm{obs}}$ and $\chi_{\mathrm{int}}$ are the observed and intrinsic EVPA, and RM is the Faraday Rotation Measure. In thermal plasma, the RM can be described by 
\begin{equation}
	\text{RM}=812\int n_e\boldsymbol{B_{\Vert}}\cdot dl	\quad \mathrm{[rad~m^{-2}]}
\end{equation}
where $n_e$ is the electron density in cm$^{-3}$ , $\boldsymbol{B_{\Vert}}$ is the magnetic field component along the line of sight, in milligauss, and $dl$ the path length in parsecs. The analysis of Faraday rotation is important to determine the intrinsic polarization angle, and the magnetic field structure of the emitting region. 

Time variability in the Faraday rotation measure was observed extensively by different authors. \citet{asada2008a} found a time variable RM in 3C273 over a period of 7 years, where the RM gradient persisted but its magnitude increased. They concluded that the variation was caused by the helical magnetic field in the jet sheath. By constraining the twist angle of the helical field (angle measured from toroidal to longitudinal component), \citet{asada2008b} found that the magnetic field that contributes to the Faraday rotation is not cospatial with the emitting region. However, the uncorrelated changes in RM and RM-corrected EVPA observed in 3C 120 by \citet{gomez2011} implied that the emitting jet and the Faraday rotating material were not physically connected in that source. The double sign reversal observed favours the RM origin at the foreground cloud instead of the magnetic tower model. \citet{gomez2008} also proposed for the source that the localized RM and the analysis of other polarization properties, indicate the jet/cloud interaction as the origin of Faraday rotation. In Mrk 421, \citet{lico2014} connected the increase in flux density from emerging components with the increase in RM, hence supporting the magnetic tower model. This is further supported by two RM sign reversals. The magnetic tower model describes an inner magnetic field with the same helicity as the accretion disc, and an outer magnetic field of opposite helicity. The change in relative contribution from the inner and outer helical magnetic field, can be related to RM sign changes. \citet{osullivan2009} suggested bends or acceleration in jets can also lead to RM sign changes. \citet{lisakov2021} then employed an oversized jet sheath model to explain the RM variability in 3C273, which should evolve slowly and be not affected much by the change in jet direction. The radial profile of the RM magnitude supports its origin in hot winds, rather than the jet sheath \citep{park2019}. The authors further explained that the dominance of negative RM without significant gradient is caused by jet and wind axis misalignment. 

The source 1633+382 is a flat spectrum Radio Quasar (FSRQ), with a redshift of $z=1.813$ \citep{hewett2010}, with black hole mass estimated between $\log{M/M_{\odot}}=7.5-9.5$ \citep{foschini2011, fan2012, chen2015, zamaninasab2014}. The parsec-scale jet is aligned at $\sim2.5^{\circ}$ to our line of sight \citep{hovatta2009, liu2010}, with superluminal motion of the jet up to $393 \pm 26\; \mathrm{\mu as/yr} \; (30.8\pm 2.0 \; c)$ detected at 15~GHz \citep{lister2021}. Intraday variability in radio spectrum was observed \citet{aller1992}, and \citet{volvach2009} found strong variability in radio, optical, and $\gamma$-ray frequencies. 

The RM in this source was observed separately in 2006 \citep{hovatta2012}, 2007 \citep{coughlan2013}, and 2008 \citep{algaba2013}. For the three studies, the observed RM were different from each other, but no variability analysis was conducted throughout the literature. Hence in this study, observations from 2004 and 2005 were added, and RM variability was analyzed together with other polarization properties.

\section{Methodology}

Multifrequency Very Long Baseline Array (VLBA) archival data was used, covering 5 epochs as shown in Table~\ref{tab:project_table}. All data sets consist of multifrequency single-epoch observations, except for year 2005, when observations from two separate close-epoch projects were combined. The data were calibrated using standard AIPS calibration techniques for polarization. Raw VLBA archive data was downloaded, then amplitude calibration, opacity correction, parallactic angle correction, fringe fitting, crosshand delay correction, and bandpass calibration were done. Imaging was carried out in DIFMAP with uniform weighting throughout the process. Simultaneous solution for instrumental polarizations (D-terms) and the source polarization of D-terms calibrator were determined using the AIPS task LPCAL. Then, the EVPAs for each epoch were calibrated using near-simultaneous observation from VLA POLCAL project data. When there is no simultaneous observation or frequencies, for each VLA frequency we first interpolated the VLA EVPA in time. Then we plotted the obtained EVPAs against $\lambda^2$ of VLA, and from there we estimated the integrated EVPAs of VLBA observing frequencies and their error.

Exception is the project BA089, for which there was no EVPA calibrator to compare with the VLA data. Hence for this epoch the EVPA was calibrated by comparing D-terms phases between two observations \citep{gomez1992}. Here we used project BO033 (7 Dec 2008). To ensure consistency, we used the D-terms calibrator 0420-014 in both projects. We then obtained the D-terms phases of all antennae, for both projects. The D-terms phase offset between project BA089 and BO033 was determined, and then averaged across antennae to recover the EVPA correction through standard polarization interferometric equation \citep{leppanen1995}. Since there was no direct comparison on the EVPA calibrator between VLBA and VLA observation, the error introduced in the final EVPA correction is directly affected by the spread of D-terms among antennae. This caused larger uncertainties in the final EVPA compared to other epochs. Calibration information is presented in Table~\ref{tab:project_table}.

After the calibration, CLEAN images of Stokes I, Q, and U were produced with two methods: the standard polarization CLEAN in Difmap, and the complex CLEAN method in Miriad. Complex CLEAN method described by \citet{pratley2016a} preserves the rotational invariance of polarization CLEANing, while reducing spurious components. We then compared the derived polarization properties. More detailed information on the comparison is discussed in the Appendix. We concluded that the standard polarization CLEANing is reliable within uncertainty, and proceeded with the standard method to maintain continuity with different studies in the literature. The errors in the CLEAN maps of Stokes I, Q, and U were then determined using the rms of residual maps. However, note that the actual error in Stokes Q and U maps needs to be further analysed to include the CLEANing error and the Dterms error. See below.

Stokes I, Q, and U CLEAN maps were convolved with the same beam size respective to the observation, as in Table~\ref{tab:project_table}. For maps of year 2004, 2005, and 2006, the beam of the lowest frequency was chosen as the convolving beam. For years 2007 \citep{coughlan2013} and 2008 \citep{algaba2013}, we used the same beam as their respective article. This is to facilitate the comparison of results.

The information about the absolute phase is lost during self-calibration while imaging, introducing an offset of the core from the map center. The position of the observed radio core will also shift with frequency, this is the well-known core shift effect \citep{lobanov1998}. The image shift, which is the vector sum of offset and core shift is required to properly align maps of all frequencies. Stokes I maps were used to obtain the image shift, and then applied to Stokes I, Q, and U maps. We used the program VIMAP developed by \citet{kim2014}, based on the cross-correlation of the optically thin region of jet \citep{walker2000, croke2008}, to obtain the image shift. 

We then used the aligned Stokes Q and U maps to construct the polarized intensity map $(p=\sqrt{Q^{2}+U^{2}})$, and the EVPA map $(\chi=\frac{1}{2}\arctan{\frac{U}{Q}})$. Rician debiasing was performed on all polarization intensity maps as in \citet{wardle1974}. Their respective noise maps were also produced. To estimate the error in Stokes Q and U maps ($\sigma_\mathrm{Q}$ and $\sigma_\mathrm{U}$), we further added the D-terms error and the CLEAN error as below \citep{hovatta2012}.
\begin{equation}
	\sigma=\sqrt{\sigma_\mathrm{rms}^{2}+\sigma_\mathrm{Dterms}^{2}+(1.5\times\sigma_\mathrm{rms})^{2}}
\end{equation}
\begin{equation}
	\sigma_\mathrm{Dterms}=\frac{0.01}{\sqrt{N_\mathrm{ant}\times N_\mathrm{IF}\times N_\mathrm{scan}}}\sqrt{I^{2}+(0.3\times I_\mathrm{peak})^{2}}
\end{equation}
where $\sigma_\mathrm{rms}$ is the error in the final residual map, the factor 0.01 is the rough estimate from the scatter of fitted Dterms, $N_\mathrm{ant}$ is the number of antennae, $N_\mathrm{IF}$ is the number of IFs, $N_\mathrm{scan}$ is the number of scans for 1633+382, $I$ is the total intensity map, and $I_\mathrm{peak}$ is the peak total intensity of the map. Then $\sigma_\mathrm{Q}$ and $\sigma_\mathrm{U}$ were propagated into other polarization properties. Note that the error from the EVPA calibration was included in the final EVPA error. To calculate the fractional polarization error, we propagated the error from $\sigma_Q$, $\sigma_U$, and the Stokes I residual map rms $\sigma_{rms}$. 

To calculate the RM, we resolved the $n\pi$ ambiguities of EVPA across frequencies. Our data show that the largest EVPA difference between the highest and lowest frequency across all epochs was only $\sim45^{\circ}$. Introducing more than one $\pi$ rotation to the EVPA can artificially smooth out the RM fitting residual, while producing extremely high RM at the core region. Indeed by introducing $n\pi$ rotation we can produce a linear $\lambda^2$ fit, but we found that the |RM| obtained this way is of the order $10^4$ rad~m$^{-2}$ or higher. RM of this order of magnitude will easily rotate the EVPA of close frequency pairs significantly. Refering to project BL137i and BG173b, such significant EVPA differences were not observed between close frequency pairs, for example 8.1 and 8.4~GHz, 4.6 and 5.1~GHz, 7.9 and 8.9~GHz; in fact the separation is merely a few degrees. Similarly to the other epochs, the $n\pi$ rotated EVPA gives an artificially high RM. Therefore, we assumed that the EVPA difference between the frequencies is only within $\pm90^{\circ}$. 
 
We then proceeded to produce the RM maps using the same-beam registered polarization maps. RM was determined by the slope of the linear fit between EVPA and $\lambda^2$. In the core region, some epochs show deviation from the linear $\lambda^2$ law, hence in these cases we either used: a) more than one linear fit, or b) the scaling relation of RM$\propto \nu^{a}$, where $a$ is the power index (see Section 3.3). In the jet region the EVPA obeyed the linear $\lambda^2$ law, hence a single linear fit was sufficient.

\begin{table*}
	\centering
	\caption{Observation epoch. [1] Observation date. [2] Project code. [3] Observation frequency. [4] Convolving beam, in minor axis FWHM, major axis FWHM, and degree eastward from north. [5] EVPA calibrator. [6] Reference antenna, same for all frequencies. [7] EVPA correction, in order of observing frequency.}
	\label{tab:project_table}
	\begin{tabular}{ccccccc} 
		\hline\hline
		Date & Project & Frequency (GHz)\textsuperscript{a} & Beam (mas, mas, $^{\circ}$) & EVPA cal. & Ref. ant. & EVPA corr. $(^{\circ})$\\
		{[1]} & {[2]} & {[3]} & {[4]} & {[5]} & {[6]} & {[7]} \\
		\hline
		1 Nov 2004 & BG152 & 15, 22, 43 & 0.50, 0.85, -12.1 & 2134+004 & LA &  $58\pm5, -51\pm6, -63\pm5$ \\
		\hline
		2 Feb 2005 & BK107f & 22, 43 & 0.60, 0.98, 19.77 & 1749+096 & LA & $15\pm4, -42\pm1$  \\
		5 Feb 2005 & BL123B & 15 & & * & * & * \\
		\hline
		6 Sep 2006 & BL137i & 8.1, 8.4, 12, 15 & 0.96, 1.35, 0.69 & 1156+295 & LA & $-3\pm3, 2\pm3, 126\pm3, 106\pm5$\\
		\hline
		26 Sep 2007 & BG173b & 4, 5, 7, 8, 12, 15 & 2.04, 3.77, -26.53 & 0851+202 & FD & $-111\pm2, 63\pm1, -106\pm4$ \\
		 & & & & & & $-29\pm1, -5\pm4, -32\pm2$ \\
		\hline
		2 Nov 2008 & BA089 & 12, 15, 22, 24 & 0.75, 0.75, 0
.00 & Dterm & LA & $55\pm6, -49\pm4, 69\pm8, -47\pm16$\\
		\hline
		\multicolumn{7}{l}{Note:} \\
		\multicolumn{7}{l}{* The calibrated archive data was obtained from MOJAVE website.}\\
		\multicolumn{7}{l}{\textsuperscript{a} For more accurate frequency value refer Table~\ref{tab:evpa_table}}
	\end{tabular}
\end{table*}

\begin{table*}
    \centering
	\caption{Core and jet polarization properties, after full calibration and corrected for $n\pi$ rotation to within 0 and 1$\pi$. The values stated below are taken at polarization intensity peak of 15~GHz. Refer to Figure~\ref{fig:result1}.}
	\label{tab:evpa_table}
	\begin{tabular}{cccccccccc}
    \hline \hline
        Project & Freq & $\chi_\mathrm{core}$ & $\chi_\mathrm{jet}$ & RM$_\mathrm{core}$ & RM$_\mathrm{jet}$ & $\chi_\mathrm{int, core}$ & $\chi_\mathrm{int, jet}$ & $m_\mathrm{core}$ & $m_\mathrm{jet}$ \\
		~	& (GHz) & $ (^{\circ})$ & $ (^{\circ})$ &  (rad~m$^{-2}$) & (rad~m$^{-2}$) & $(^{\circ})$ & $(^{\circ})$ & (\%) & (\%) \\ \hline
        BG152 & 15.285 & $20\pm5$ & $63\pm8$ & $-1496\pm923$ & * & $49\pm14$ & * & $2.25\pm0.08$ & $14.7\pm2.2$ \\
        ~ & 22.235 & $22\pm8$ & * & $\sim 0^{22,15}$ & ~ & ~ & ~ & $4.10\pm0.12$ & * \\
        ~ & 43.135 & $50\pm7$ & * & $-3656\pm1400^{43,22}$ & ~ & $60\pm10^{43,22}$ & ~ & $5.39\pm0.12$ & * \\ \hline
        BL123B & 15.365 & $44\pm1$ & $59\pm3$ & $-635\pm299$ & * & $56\pm6$ & * & $6.11\pm0.05$ & $15.3\pm1.2$ \\
        BK107f & 22.233 & $46\pm1$ & $68\pm4$ & $\sim 0^{22,15}$ & ~ & ~ & ~ & $6.39\pm0.04$ & $14.0\pm1.4$ \\
        ~ & 43.217 & $61\pm2$ & * & $-1824\pm184^{43,22}$ & ~ & $65\pm4^{43,22}$ & ~ & $4.44\pm0.10$ & * \\ \hline
        BL137i & 8.108 & $34\pm4$ & $52\pm5$ & $-601\pm120$ & $-421\pm127$ & $81\pm8$ & $84\pm8$ & $1.81\pm0.12$ & $11.2\pm0.8$ \\
        ~ & 8.428 & $39\pm5$ & $52\pm5$ & ~ & ~ & ~ & ~ & $1.71\pm0.10$ & $11.4\pm0.7$ \\
        ~ & 12.123 & $57\pm5$ & $68\pm5$ & ~ & ~ & ~ & ~ & $1.41\pm0.09$ & $12.8\pm0.9$ \\
        ~ & 15.365 & $74\pm7$ & $78\pm8$ & ~ & ~ & ~ & ~ & $1.44\pm0.09$ & $13.5\pm1.1$ \\ \hline
        BG173b & 4.612 & $51\pm4$ & $68\pm5$ & $390\pm82$ & $45\pm15$ & $24\pm5$ & $58\pm2$ & $1.89\pm0.10$ & $8.2\pm0.3$ \\
        ~ & 5.092 & $54\pm2$ & $66\pm2$ & ~ & ~ & ~ & ~ & $2.03\pm0.09$ & $8.7\pm0.3$ \\
        ~ & 7.916 & $54\pm6$ & $61\pm5$ &  $-69\pm103^{5,4}$ & ~ & ~ & ~ & $1.48\pm0.09$ & $9.0\pm0.3$ \\
        ~ & 8.883 & $50\pm3$ & $61\pm2$ & ~ & ~ & ~ & ~ & $1.42\pm0.09$ & $9.4\pm0.4$ \\
        ~ & 12.939 & $34\pm5$ & $61\pm5$ & ~ & ~ & ~ & ~ & $1.38\pm0.07$ & $9.7\pm0.4$ \\
        ~ & 15.383 & $32\pm3$ & $57\pm3$ & ~ & ~ & ~ & ~ & $1.37\pm0.08$ & $10.1\pm0.6$ \\ \hline
        BA089 & 12.039 & $65\pm8$ & $63\pm8$ & $355\pm668$ & $79\pm499$ & $52\pm20$ & $60\pm13$ & $1.08\pm0.10$ & $20.7\pm1.7$ \\
        ~ & 15.383 & $60\pm8$ & $62\pm7$ & ~ & ~ & ~ & ~ & $0.74\pm0.09$ & $19.7\pm1.7$ \\
        ~ & 21.775 & * & $62\pm12$ & ~ & ~ & ~ & ~ & * & $18.5\pm2.6$ \\
        ~ & 23.998 & $55\pm21$ & $59\pm19$ & ~ & ~ & ~ & ~ & $0.45\pm0.09$ & $21.0\pm2.4$ \\ \hline
	\multicolumn{7}{l}{Note:} \\
	\multicolumn{7}{l}{* No polarization intensity more than $3\sigma$ observed.}\\
	\multicolumn{7}{l}{** Error quoted according to Monte Carlo simulation.}\\
	\multicolumn{7}{l}{Value with superscript \textsuperscript{a,b} is obtained through fitting the fitting of frequencies a~GHz and b~GHz.}
    \end{tabular}
\end{table*}

\section{Results}

\begin{figure*}
	\includegraphics[scale=0.8]{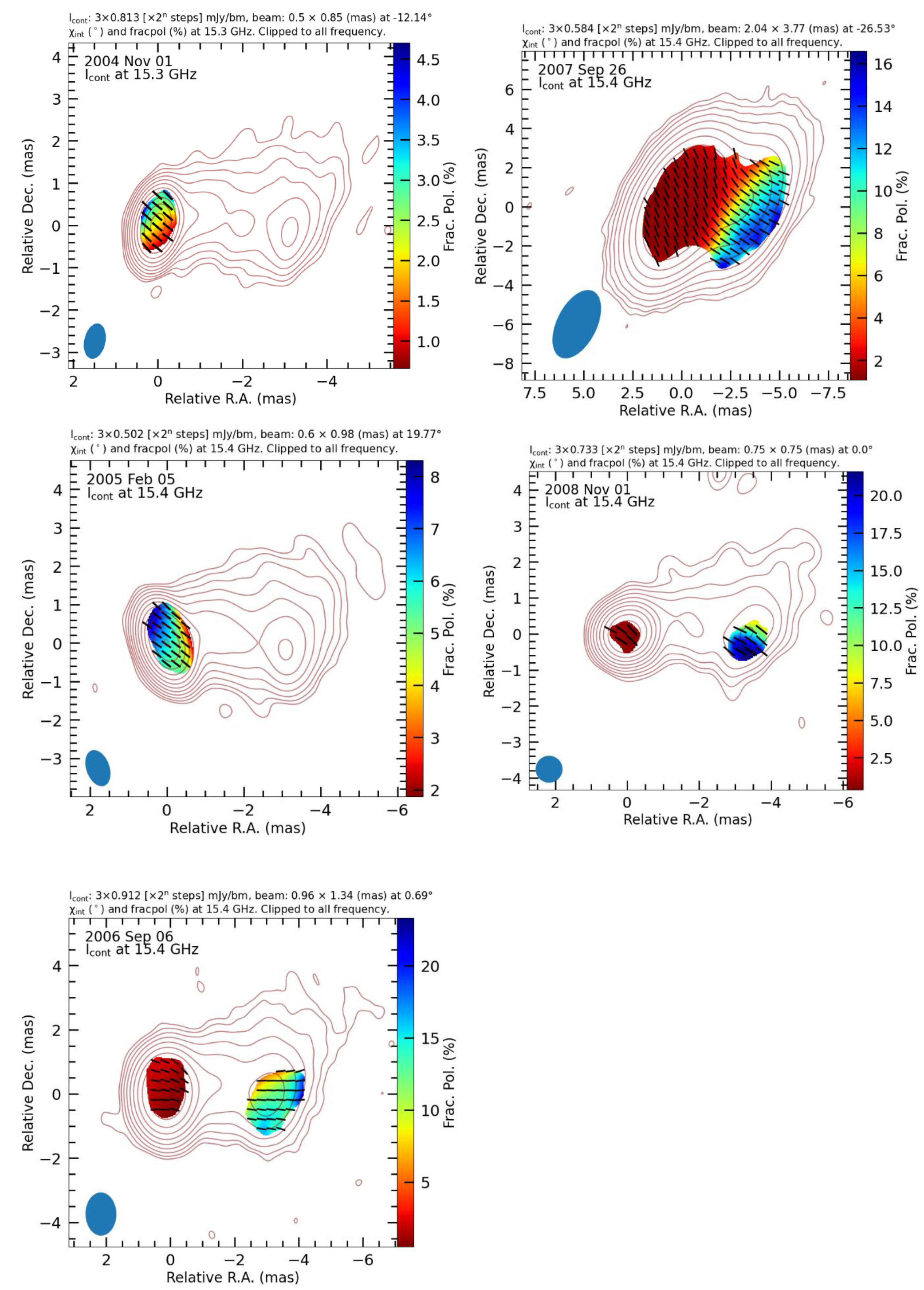}
	\caption{Maps of 1633+382 at various epochs discussed. The contour levels indicate the total intensity, starting at $3\sigma$ level, increases at steps of $2^n$. Colour scale indicate the fractional polarization. Tick mark of uniform length represents $\chi_\mathrm{int}$. The solid blue ellipse on the bottom left corner indicate the beamsize of the map, For fractional polarization and EVPA maps of other frequencies, see Appendix C.}
	\label{fig:pa0}
\end{figure*}

\begin{figure*}
	\includegraphics[scale=0.8]{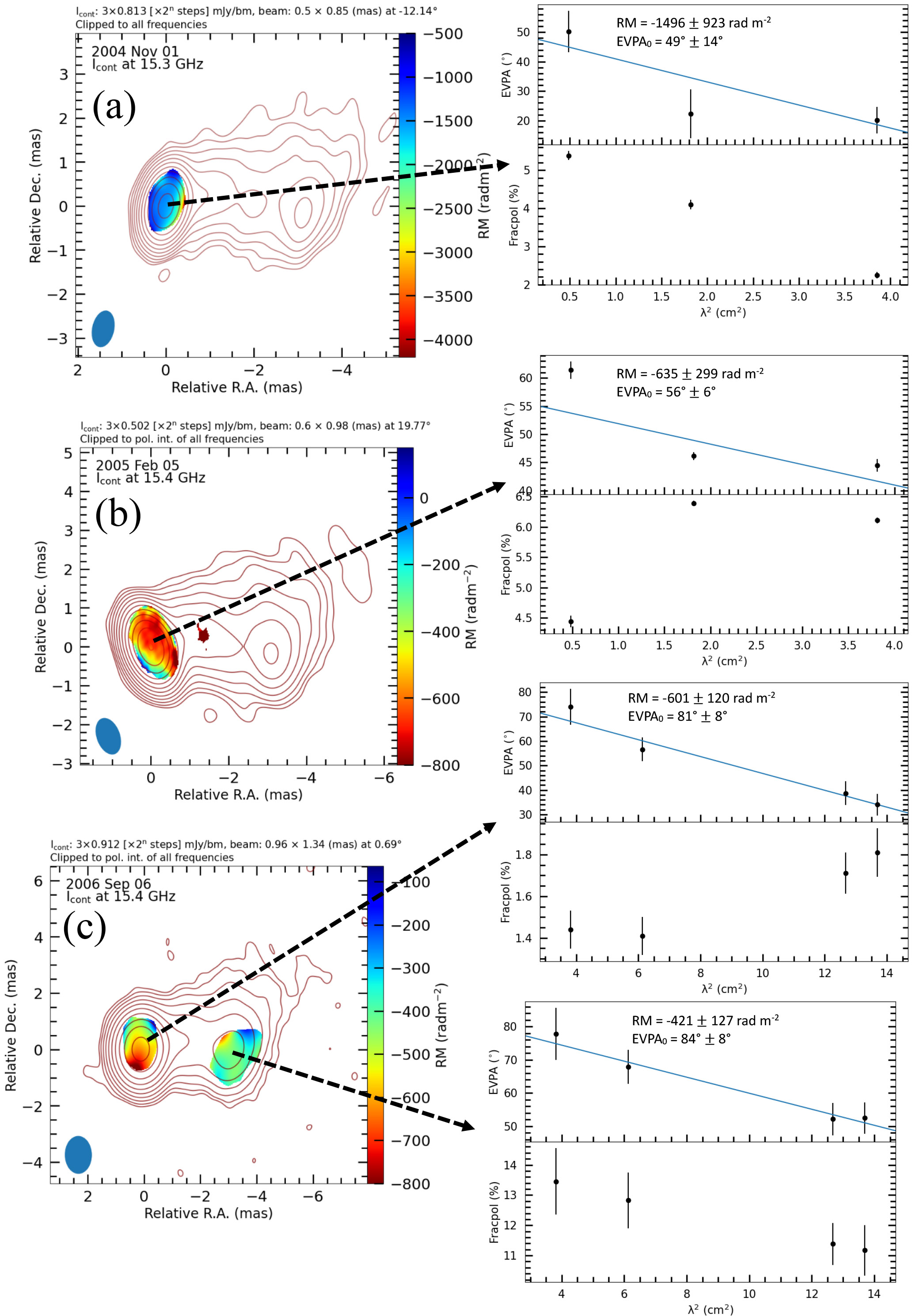}
	\caption{RM maps of all epochs. RM colour map overlayed on total intensity contour, arranged in timely order. The total intensity contours were plotted in levels $3\sigma \times 2^n$ steps, where the $\sigma$ is the residual rms. The RM fit (shown as blue or orange colour line) and fractional polarization at the indicated position were shown in the plot. In \textbf{(d)} the colour scaling is different in the core and jet. In the core the RM value fitted with the highest four frequencies was shown as colour scale, the RM obtained from the lowest two frequencies was only quoted at the right of the panel. In the jet region all six frequencies were used. The black curved line indicates the separation of core and jet region. In \textbf{(e)} the core region RM fit excluded 22~GHz, because we do not detect significant 22GHz polarization in core. The vertical black line at the jet region indicates the slice position. }
	\label{fig:result1}
\end{figure*}

\renewcommand{\thefigure}{\arabic{figure} (Cont.)}
\addtocounter{figure}{-1}

\begin{figure*}
	\includegraphics[scale=0.8]{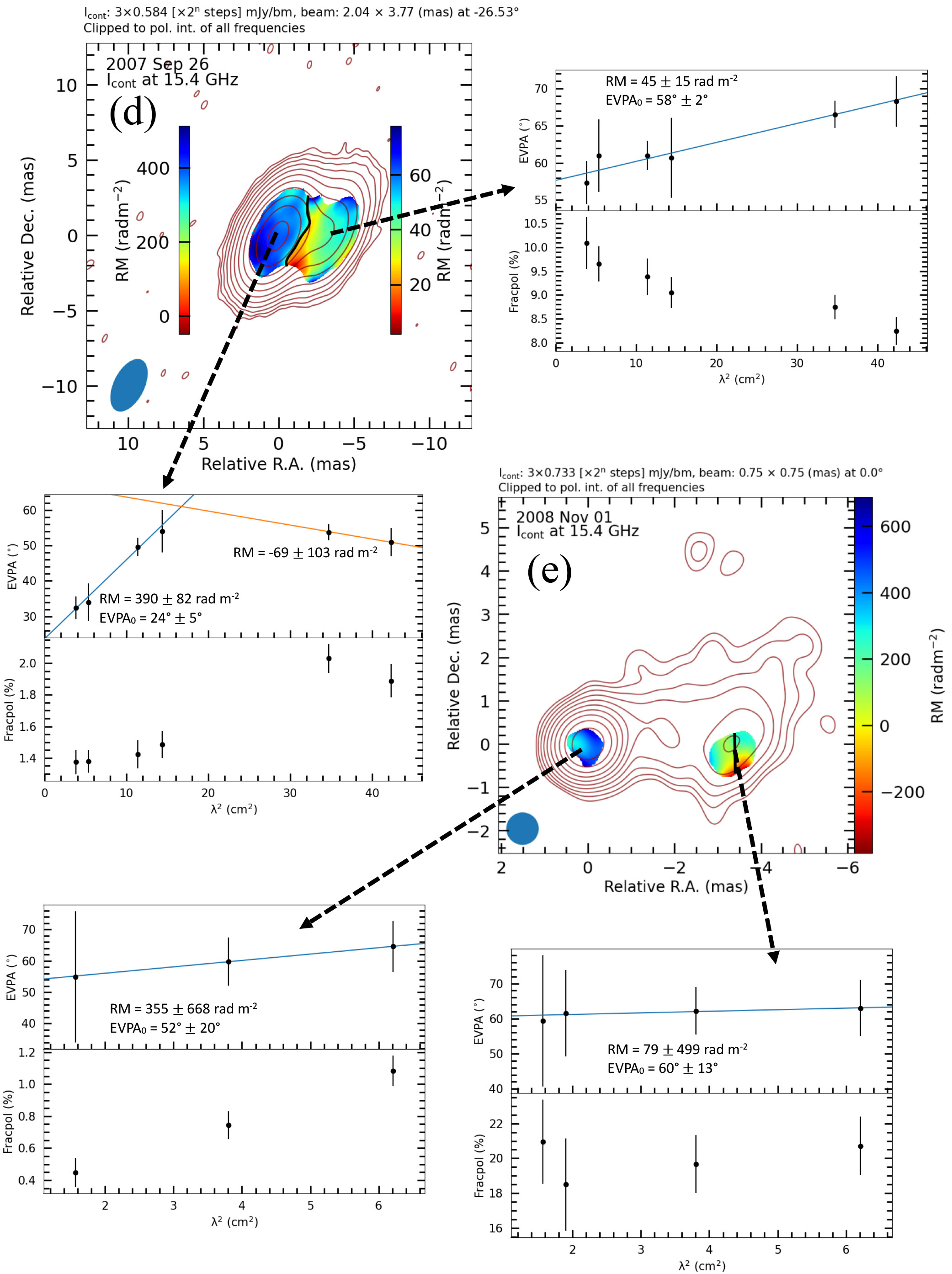}
	\label{fig:result2}
	\caption{ }
\end{figure*}

\renewcommand{\thefigure}{\arabic{figure}}

\subsection{Polarization properties}

We obtained total intensity, polarization intensity, fractional polarization, and Faraday corrected EVPA maps for five epochs.  In general, we observed significant polarization in two regions of the source: in the core region, and in the standing knot of the jet located at around 3.5~mas from the core.  In both cases, the region with polarized emission extends only 1 or 2 beamwidths. The fractional polarization and Faraday-corrected EVPA maps are shown in Figure~\ref{fig:pa0}. We observe that both the core and jet fractional polarization show a gradient. Because of the unresolved nature of the core, the gradient at the core does not allow for further interpretation. The gradient at jet region will be discussed at Section 4.5. The Faraday corrected EVPA was oblique to the jet direction, both in the core and jet region. In the jet region, the Faraday corrected EVPA was roughly perpendicular to the fractional polarization gradient, for years 2006, 2007, and 2008. 

\subsection{General RM maps}
RM maps for all epochs are shown in Figure~\ref{fig:result1}. At the core region, we observe an increase in |RM| with frequency in three epochs. This deviation from the linear $\lambda^2$ law may indicate opacity effects. The exceptions are 2006 and 2008, where in 2006 the EVPAs obeyed the linear $\lambda^2$ law, and in 2008 the EVPA calibration error is too large. Because of: a) sparse frequency sampling, or b) the overall frequency span is too small, we were not able to definitely rule out internal Faraday rotation. We assumed an external Faraday screen because of slow variability in 43-22~GHz RM. In the jet region the linear $\lambda^2$ law was observed for 2006, 2007 and 2008. We also observed the inverse depolarization in the core region, except in 2004 where depolarization increased toward low frequency. Here we specify further comments on the years 2006, 2007, and 2008. 

\subsection{Sept 2006}

The core EVPA scales linearly with $\lambda^2$. The core region shows higher |RM| than the jet region. The core RM tends to have a higher magnitude northward, but because of the unresolved nature of core region we do not consider the presence of gradient. See more details in Section 4.3.

We further investigated the possibility of deviation from the linear $\lambda^2$ law. From RM$\propto \nu^a$ \citep{jorstad2007a}, we used |RM| $=A\times \nu^a + d$ for generality. Then we derived the scaling relation for $\chi_{obs}=A\lambda^{-a+2} + \lambda^{2}d + c$. We note that this formula is only meaningful for the determination of the value $a$, and the other parameters imply the physical condition at different frequencies. Two additional constraints are applied to the above relation, i) the observed RM is of negative sign; ii) the magnitude |RM| decreases with longer wavelength. These constraints indicate that $A<0$ , $d<0$, and $a\geqslant0$. We then used the Markov Chain Monte Carlo method to find the posterior distribution of parameters $A$, $a$, $d$ and $c$. The posterior distribution of $a$ peaks strongly at zero, hence a single linear RM fit can explain the observed EVPA.

\subsection{Sept 2007}
In this epoch, the core RM sign also changes from positive to negative going toward lower frequency, with a decrease in |RM|. As shown in Figure~\ref{fig:result1} (d), the core EVPA does not scale linearly with $\lambda^{2}$. The increase in RM with shorter wavelength is expected from the scaling relation RM$\propto \nu^{a}$. Using the two core RM fitted in Figure~\ref{fig:result1} (d), we obtained the power index $a\sim1.48\pm1.40$.

In the jet region, RM scales linearly with $\lambda^{2}$, which is consistent with the optically thin regime. Spatially, the RM obtained with 4.6 and 5.1~GHz exhibited sign change from negative to positive going from the core to the jet region. 

\subsection{Nov 2008}
The core region shows a higher |RM| than the jet, while having a similar positive sign. The large uncertainty at 24~GHz, and the fact that only three frequencies are observed, does not permit the observation of EVPA scaling with $\lambda^2$. A slice was taken at the jet region, from south to north, as shown in Figure~\ref{fig:result1} (e). A RM gradient was observed in this region, is consistent with \citet{algaba2013}. The fractional polarization of 12~GHz shows signs of depolarization at the southern $\sim$0.3 mas of the slice, and the peak fractional polarization coincides with zero |RM| value. The depolarization effect is weaker for 15~GHz, and not observed in higher frequencies. Reaching the center of the knot, all frequencies show similar magnitude of fractional polarization, even with the increasing |RM|. Going northward, all frequencies show low fractional polarization, without significant difference in depolarization. 

\begin{figure}
	\includegraphics[width=\columnwidth]{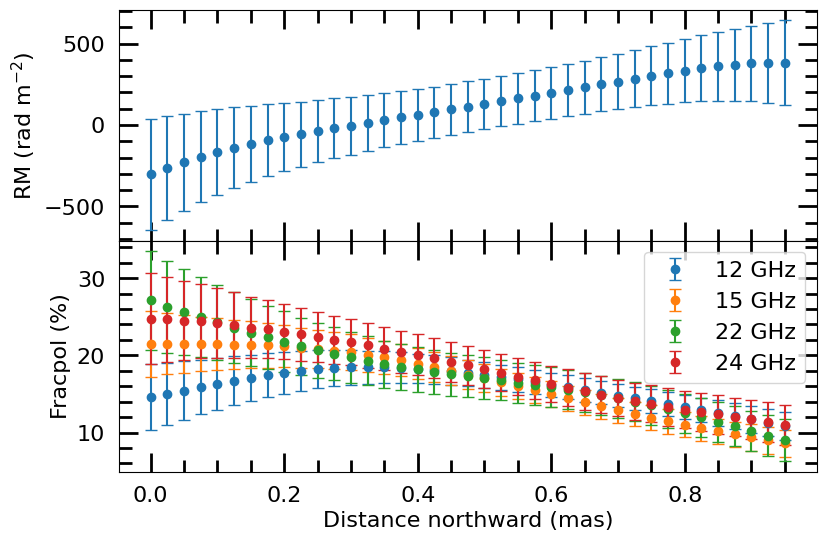}
    \caption{Slice took from south to north of jet region, year 2008. Both RM and fractional polarization profile shows transverse gradient. The error bar in RM plot excludes the EVPA calibration error, as it is correlated spatially hence irrelevant in the determination of RM gradient \citep{mahmud2013}. Note that the beam size spans 0.75~mas.}
    \label{fig:slice2008}
\end{figure}

\section{Discussion}
\subsection{High frequency core RM}
The 2004 and 2005 observations are spaced by 3 months, the shortest within the five epochs. In this period, the core RM sign remains negative. The apparent 43-22~GHz |RM| decrease in this period is within $1\sigma$ of the error, and RM for 22-15~GHz RM remains at zero. The $\chi_{\mathrm{int}}$ obtained using 43-22~GHz are consistent within $1\sigma$. Hence the core RM sign, magnitude, and $\chi_{\mathrm{int}}$, is consistent within 3 months, at least within the errors of our observation. Hence the Faraday screen in the core region could be external to the jet \citep{hovatta2012}.

Following the assumption of external Faraday screen, \citet{jorstad2007a} derived |RM|$\propto d^{-a}$, where $d$ is the distance from central engine. Then the core distance from central engine depends on observing frequency $\nu$, as derived by \citet{lobanov1998} as $d_\mathrm{core,\nu}\propto \nu^{1/k_r}$. The value of $k_r\sim1$ in 1633+382 is obtained by \citet{algaba2012}, hence from the relations above we can write |RM$_{\mathrm{core},\nu}$|$\sim \nu^{a}$. Using this relation, we extrapolated 2006 RM to 43-22~GHz to investigate the longer time scale change. If we use the power index of year 2007 $a \sim 1.48$ (since in 2006 we could not obtain a power index) to extrapolate to 43-22~GHz, we obtained RM$\sim -3000$~rad~m$^{-2}$. RM in 2006 remains negative, similar to 2004 and 2005.  

Here we provide an interpretation for 43-22~GHz observation (15~GHz in Section 4.4). The 43-22~GHz |RM| shows remarkably slow variability in year 2004, 2005, and possibly also 2006. However similar slow variability is not observed in the 43~GHz EVPA. We obtained from Boston University Blazars\footnote{\url{https://www.bu.edu/blazars/VLBA_GLAST/1633.html}} monitoring program 2007 to 2008 data, the 43~GHz polarized intensity and EVPA varies rapidly on timescale of 1-2 months. We can assume a single emitting component with an external Faraday screen, dominating the 43-22~GHz core flux. In this case both the $\chi_{\mathrm{int,43-22GHz}}$ and 43~GHz EVPA should show similar variability, but their variability should be uncorrelated to that of RM. Indeed from 2004 and 2005 both $\chi_{\mathrm{int,43-22GHz}}$ and 43~GHz EVPA show similar increment (within error of $\sim$10$^{\circ}$). This agrees with external Faraday screen interpretation in previous paragraph. This correlation is not guaranteed if there are multiple components, each dominating the polarized flux at different times. Unfortunately we do not have 43~GHz data in 2006 to estimate $\chi_{\mathrm{int,43-22GHz}}$. To confidently determine the variability of RM, $\chi_\mathrm{int}$, and EVPA, denser frequency and time sampling are imperative.

\subsection{Equipartition condition}
From the relation $\mathrm{RM}\propto \nu^{a}$ \citep{jorstad2007a,lobanov1998}, the unknown parameters are the power index $m$ of magnetic field ($B\propto d^{-m}$), $n$ of electron density ($n_{e}\propto d^{-n}$), and $k_r$ related to the core shift effect. We assume here a conical jet and toroidal magnetic field \citep{jorstad2007a}. The scaling of magnetic field and electron density here refer to the quantity in the jet sheath, and $k_r$ refers to the jet emitting region. With these conditions the value of $a$ is described by $a=(m+n-1)/k_r$, hence $a=m+n-1$. Then the value of $a$ reflects the scaling relation of jet sheath. If the jet sheath follows equipartition, then the reasonable estimates are $a=2$, $m=1$, and $n=2$. In all five epochs the only estimate of $a$ is in year 2007, where $a=1.48\pm1.40$, approximately $a\sim2$ although with large uncertainty. This may indicate that the jet sheath is in equipartition condition.

\subsection{Unresolved nature of core}
We further investigated the possibility of resolving the RM structure in the core. \citet{mahmud2013} simulated transverse RM gradient of a given width, and then convolved with a beam of size 20 times larger. Their result showed that the gradient remained present, but the magnitude diminished. Here we intended to constrain the beam-to-core size ratio, to identify any underlying structure. We used the total intensity (Stokes I) map as the baseline of resolving. From the Stokes I CLEAN image, we identified the region of significant emission (flux density larger than 3 times the residual rms), and the background noise region (flux density smaller than 3 times the residual rms). In the significant emission region, the brightest pixel of central unresolved core was used as the origin. Then the pixels of radial distance larger than 1.5 times the beam pattern is replaced with a simulated background noise. This will remove the jet emission while including any structure at the core. Then we fitted a 2D gaussian to the map, and obtained their FWHM for major and minor axes. We compared the FWHMs with the CLEAN beam FWHMs. This is based on a few arguments: a) Assume that the true core structure in perfect resolution is a 2D Gaussian, then convolving a \textit{resolved} Gaussian core structure with any smaller beam will only lead to a larger FWHM. This means that the beam and the component FWHM are added in quadrature. b) Assume the true core structure is other than 2D Gaussian, then the convolved image will deviate from a 2D Gaussian and lead to a bad fit. Our result shows that the fitted FWHMs and beam FWHMs are consistent to within 1\%, showing that the true core structure is a $\delta$-function, unresolved in this case. Hence we found that the core size, even if present, is at most $\sim$0.1 times the beam size.

Using the resolution criteria in \citet{kovalev2005}, with the longest baseline visibilities projected north-south (to resolve the transverse structure), we determined that the Stokes I and the polarized intensity core are unresolved. To further constrain the size, we applied the maximum theoretical over-resolution power \citep{marti2012}, and determined the 15~GHz Stokes I core can range between 0-0.015~mas ($\sim$2\% of 15~GHz beam FWHM). The core polarized intensity size has a much larger uncertainty range, 0-0.2~mas.

\begin{figure}
	\includegraphics[width=\columnwidth]{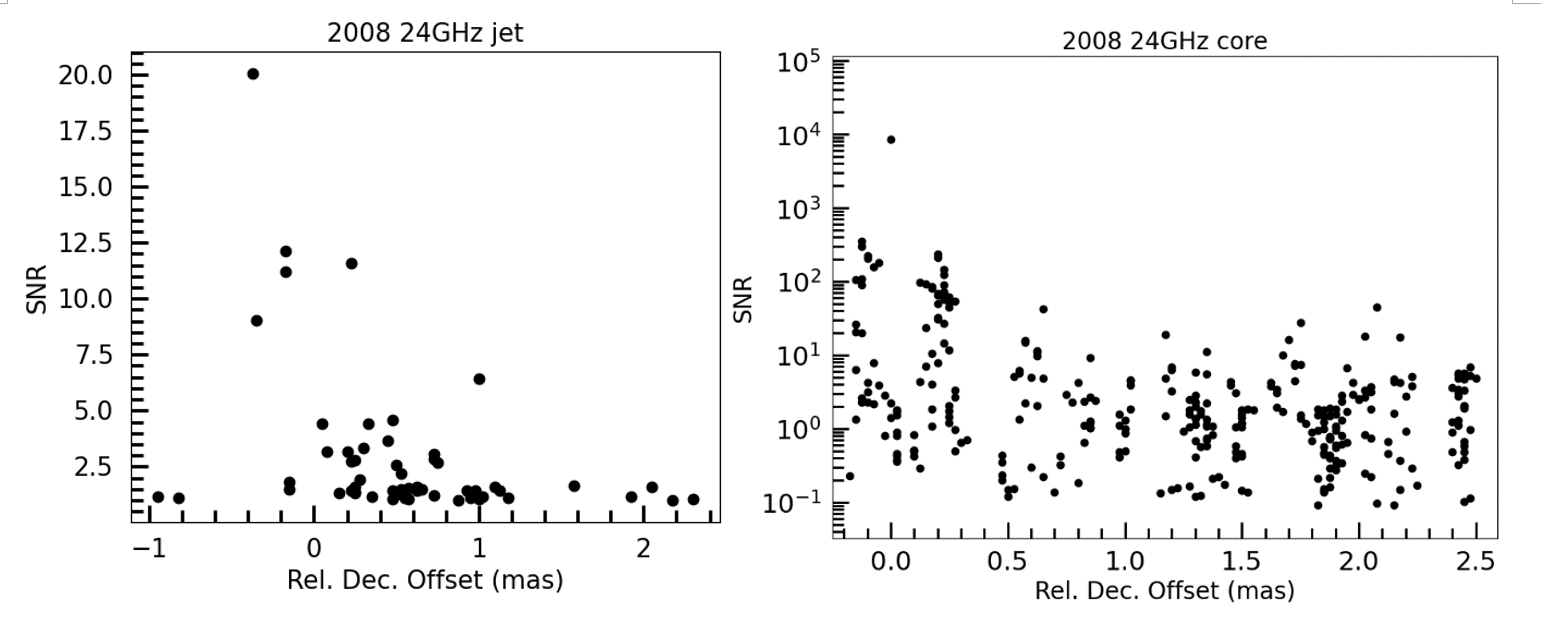}
     \caption{The declination offset of all CLEAN components larger than $2.5\sigma$. The jet is pointing westward, hence we measure declination offset to determine if the jet transverse structure is resolved. Refer to text for the meaning of SNR. Note that the plot for the jet is relative to 3.5~mas west from the image center, and the plot for the core is in logarithmic scale.}
	\label{fig:unresolve}
\end{figure}

To support this claim, we further examined the distribution of CLEAN components in the observable region. We assumed that the CLEAN delta components comprise of "true" CLEAN components, and "noise" components randomly distributed in flux density. The noise components will make up the majority of all components, hence we recursively fit the Gaussian distribution onto all CLEAN components, and isolated out the true components. We then obtained the sigma of the Gaussian fitting, then for all components, we plotted the ratio of flux density over sigma, against the declination offset from the map center. The ratio is named "SNR" in Figure~\ref{fig:unresolve}. As shown in there, there is a single extremely strong component at more than 2000 SNR, and this single component will dominate all the flux density observed at the core. This is in contrast to the jet region, where multiple components of comparable magnitude exist across a distance larger than the beam FWHM. Furthermore, the 2D Gaussian fit on the isolated jet's knot produced a FWHM twice the beam FWHM. In all epochs, the core region is dominated by a single component, which means the core is likely unresolved. Hence we avoided interpreting the polarization spatial structure, as it may arise from artifacts.

\subsection{Core RM changes}
In the core region, for both year 2004 and 2005 there is no EVPA rotation between 15~GHz and 22~GHz, hence we do not expect there will be significant rotation at lower frequency. However, in year 2006, rotation of more than 45$^{\circ}$ was observed, and this was accompanied by a 15~GHz fractional polarization decrease from 6.11\% (2005) to 1.44\% (2006). The 15~GHz low fractional polarization in year 2006 cannot be explained by RM gradient, since this substantial depolarization requires a RM gradient >2500~rad~m$^{-2}$ \citep{zavala2004}. Hence, we attribute these changes to the ejection of a new component C10 on 26 May 2005 as identified by \citet{lister2009}), observed at 15~GHz. 

\begin{figure}
	\includegraphics[width=\columnwidth]{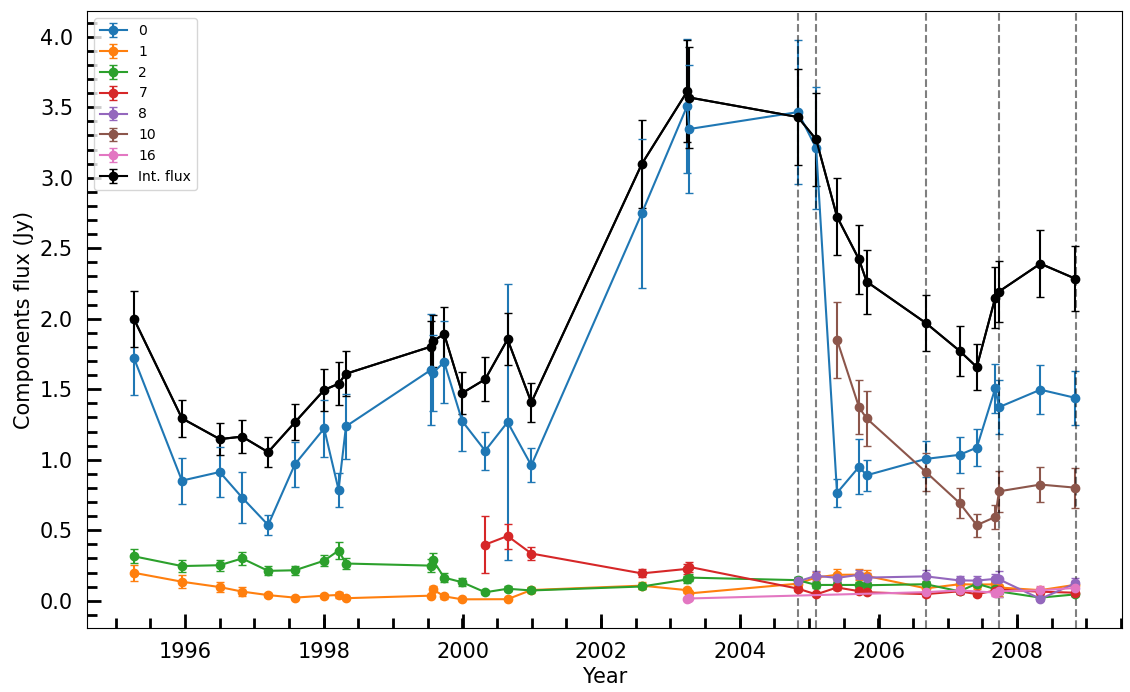}
     \caption{Components and integrated core flux density against time. The component number follows \citet{lister2009}. The vertical lines indicate the 5 epochs in this study.}
	\label{fig:coreflux}
\end{figure}

To further investigate this possibility, we obtained the MOJAVE 15~GHz MODELFIT components from \citet{lister2009}. For epochs not present in their data set, we used MODELFIT in DIFMAP to obtain the circular components. These epochs are 26 May 2005, 6 Sept 2006, 26 Sept 2007, and 2 Nov 2008. We identified three components that are located within the 1.0~mas core region. They are component numbers 0 (core), 10, and 7, which we will denote here as C0, C10, and C7. To compare the contribution of the components to the observed core flux density, we used the 15~GHz CLEAN delta components from MOJAVE. We calculated the core integrated flux density as the sum of all CLEAN delta components within the radius 1.0~mas from core (1.0~mas is the typical beam size of VLBA at 15~GHz). We then assumed the error to be 10\% of the total integrated flux density from the whole source, the typical self calibration error. The error in MODELFIT components was calculated according to \citet{lee2008} and \citet{fomalont1999}. We plotted the time evolution of the core integrated flux density together with the MODELFIT components, as in Figure~\ref{fig:coreflux}. As shown in Figure~\ref{fig:coreflux}, the two components that dominate the core emission are C0 and C10. 

\begin{figure}
	\includegraphics[width=\columnwidth]{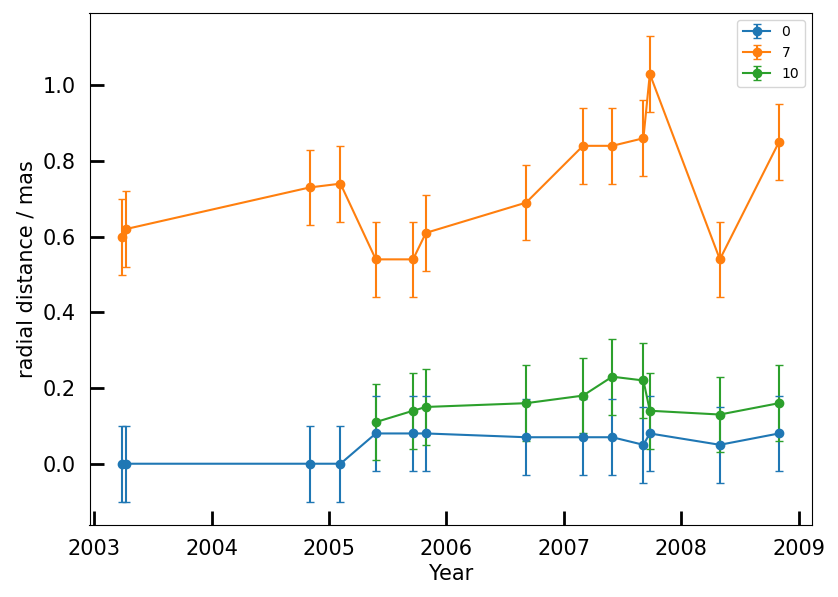}
     \caption{Motion of component C0, C7, and C10 over time. Plotted with data from \citet{lister2009}.}
	\label{fig:compmotion}
\end{figure}

Component C10 is only detected from middle of 2005 onward, and it shows a slow but linear trajectory relative to the core position (C0). The component motion of C0, C7, and C10 are shown in Figure~\ref{fig:compmotion}. Uncertainty in position was estimated using 20\% of beam size \citep{lister2009,homan2002}. We fitted the components motion and found that the predicted position on our observation date in 2008 is only $0.27\pm0.04$~mas in radial distance relative to the core position (component C0). Hence the ejected new component will not be resolved. As shown in Figure~\ref{fig:coreflux}, in year 2005 the component C10 contributed the majority of total flux density, but it decayed to a similar level on our observation date of year 2006. Then for 2007 and 2008 the core component again dominated the total flux density. Following the trend of C10, it is reasonable to suspect that the increase of core flux density starting from around 2001 is caused by the coming C10, where the core component maintains a rather stable flux density throughout the year. We extrapolated the position of C10 to year 2001, and the radial distance from core C0 is estimated to be $0.05\pm0.06$ mas, hence C10 should not be too far upstream of the jet at the time of flux density increase.

We speculate that the RM at both years 2004 and 2005 is dominated by C10, which at these dates had not moved out from the optically thick region yet. Unfortunately we did not have sufficient time coverage and frequency sampling around 15~GHz to test this hypothesis, but we observed a possible increase in RM with higher frequency. The extrapolated RM of the year 2006 to 43-22~GHz showed the same negative sign and rather similar magnitude with the previous two years. In 2006, the component C10 moved into an optically thin region and contributed to the majority of observed Faraday rotation. This can be seen from the $\lambda^2$ dependence. Then going into year 2007, the C10 flux density decays to around half of C0, hence the core observed RM follows the emission of C0. This is further supported by the 15~GHz RM sign change to positive in 2007, with the decay of C10. The year 2007 also showed a broken RM fitting between 15~GHz and 4~GHz, hence possibly probing at different regions/components at the unresolved core.

\subsection{Jet fractional polarization}
We observed that, in all epochs, the fractional polarization profile shows some degree of transverse gradient. The fractional polarization increases towards at least one of the jet edges, as in Figure~\ref{fig:pa0}. This could be caused by several different scenarios, such as a shear/compression during jet-medium interaction, or the presence of a helical magnetic field. In addition, the fractional polarization gradient appears at the apparent jet bending position, at about 3.5~mas from core.

\begin{figure}
	\includegraphics[width=\columnwidth]{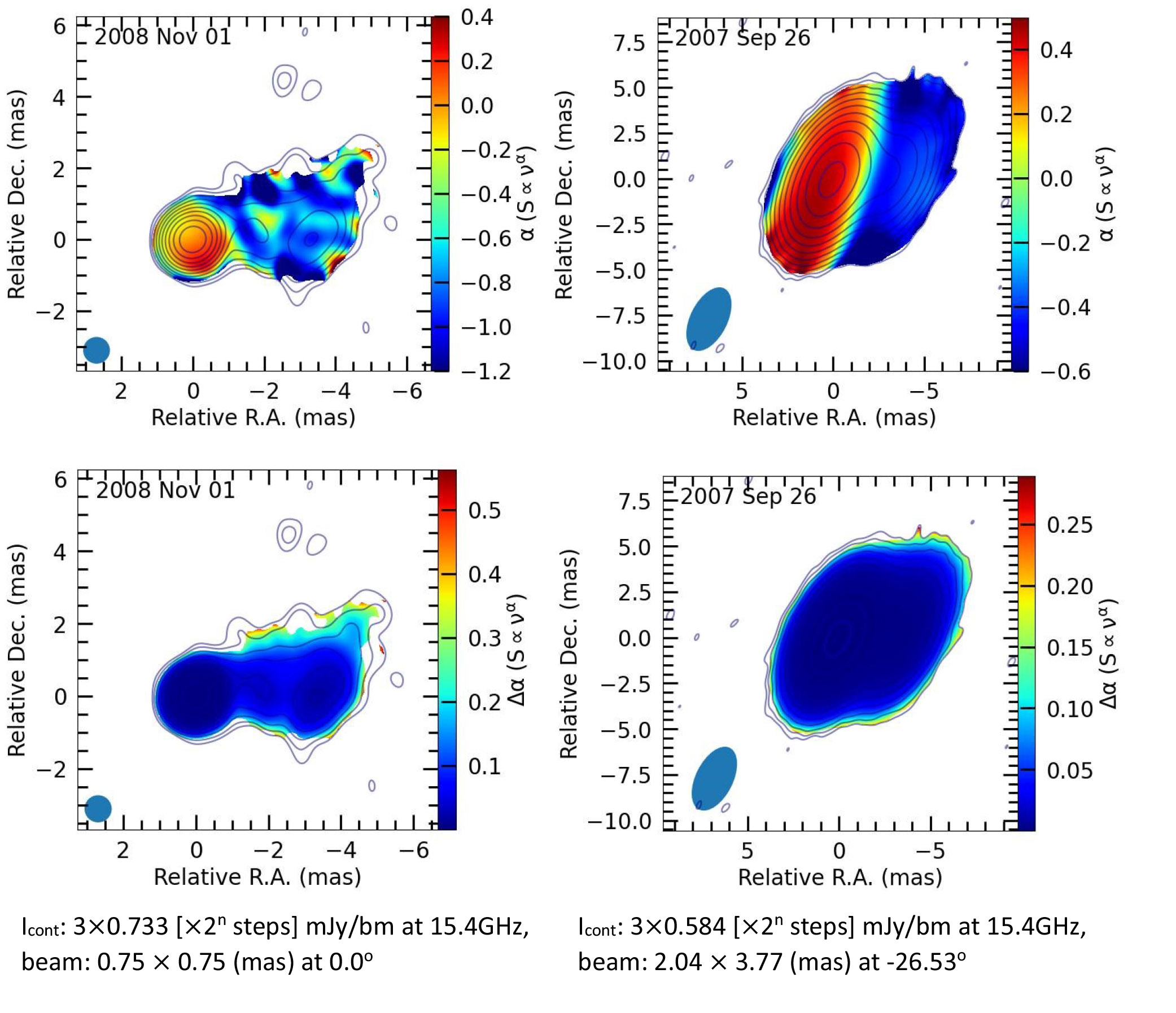}
	\caption{Spectral index map of the year 2007 (right) and 2008 (left). The top panels show the spectral index fitted from all observing frequencies, bottom panels show the error in the fitting.}
	\label{fig:spixmap}
\end{figure}

\citet{algaba2019} found that the jet trajectory follows a sinusoidal pattern, with a precession model providing a better fit. However they do not have sufficient data to rule out helical structure, instabilities, or jet-medium interaction. Jet compression by ambient medium can increase synchrotron self-absorption, which can also lead to an inverted or flat spectrum \citep{hovatta2014,mimica2009}. Other indications of jet medium interaction are the inverted spectrum of free-free absorption \citep{park2023,kino2021}, and spectral index gradient \citep{osullivan2009}. For this purpose, we obtained the spectral index map of 2006\footnote{\url{https://www.cv.nrao.edu/MOJAVE/spmaps/1633+382.2006_09_06_alpha_paper_gray.png}}, 2007, and 2008. For 2007 and 2008, we used the same-beam registered Stokes I maps of all frequencies, and fitted the relation $S\propto \nu^{\alpha}$ to find the spectral index $\alpha$. The error was determined through the conventional covariance matrix of the fit. The spectral index maps and their respective error are shown in Figure~\ref{fig:spixmap}. For three years, we did not observe any spectral index gradient at the jet region, nor an inverted spectrum at the western edges of jets. Hence we considered jet-medium interaction to be less likely, although we cannot rule out the possibility. 

In Figure~\ref{fig:slice2008}, we observe depolarization at the jet southern edge. The depolarization is only significant on 15 and 12~GHz. Beam depolarization by RM gradient cannot explain this, as a similar level of depolarization is not observed at the northern part of the slice. Hence the apparent RM gradient cannot be confirmed through beam depolarization. In addition, the RM gradient is smaller than three beamwidths, hence not fully resolved \citep{zavala2003}.

\subsection{Origin of jet Faraday screen}
For the jet region, we postulate an external Faraday screen. Here we examine the possibility of a foreground cloud using variability arguments. The RM change in the jet region does not correlate with the RM change in the core region. Hence we considered that if the Faraday rotation is to be caused by the foreground cloud, the size would be around $\sim$2 mas, or maximum $\sim$17 pc. Hence using a toy model we estimate the fluctuation in physical parameters to reproduce the observed RM change. The major RM changes happened between 2006 and 2007, with a change in magnitude |RM| of about 380~rad~m$^{-2}$, with a sign change from positive to negative. The magnetic field strength obtained by \citet{algaba2018} through synchrotron self-absorption analysis is $B_{SSA}=0.07$~mG. We will use this value as the upper limit for magnetic field strength as the strength is expected to decrease downstream toward the jet region that we observe. For electron density, the fiducial estimate we used is the value at Bondi radius, in M87\footnote{We note that the initial approximation of $n_e$ is based on a Fanaroff-Riley I (FRI) object, while our 1633+382 is an FRII which often showed signatures of mergers \citep{ramos2012,baldi2008}.} \citep{russell2015}. We then extrapolated to the jet region using the scaling relation of electron density with distance along the jet. The Bondi radius can be obtained using $r_{B}=2GM/c_{s}^{2}$, where the sound speed is $c_{s}=\sqrt{\gamma k_{B} T/\mu m_{p}}$ \citep{bondi1952}, and $M$ is the black hole mass $M=\num{1.0e9}~\mathrm{M_{\odot}}$ \citep{zamaninasab2014}, $\gamma=5/3$ is the adiabatic index of the accreting gas, $\mu=0.6$ is the mean molecular weight, $m_p$ is the proton mass, and $T=\num{3.5e6}~\mathrm{K}$ \citep{bednarek1995}. The estimated $r_B \sim110$~pc or $0.6$~mas projected, where the core region lies within the Bondi radius. Then we assumed the electron density $n_e$ at Bondi radius is 0.5~cm$^{-3}$ \citep{russell2015}, the scaling of $n_e\propto d^{-n}$, where $n=2$. Hence at jet region, the electron density would be around ~0.014~cm$^{-3}$. 

Then we estimated the fluctuation in electron density/magnetic field needed to reproduce the |RM| change of 380~rad~m$^{-2}$. First, we assumed that both the path length $l=0.1$~kpc, and magnetic field of $B=0.07$~mG are constant, with only the electron density is changing (from initial of $n_e=0.014$~cm$^{-3}$ as obtained). Then we would need a change in electron density of $\Delta n_e =0.066$~cm$^{-3}$ to reproduce the observed |RM| change from year 2006 to 2007. The change would be significantly larger than the $n_e$ expected at the region. If in turn, we assume the path length and electron density to be constant, then the magnetic field strength must fluctuate by $\sim0.33$~mG (from the initial of 0.07~mG).

Another approach to estimate the electron density and magnetic field fluctuation is to assume equipartition condition. Here we assumed that the total internal energy is equal to the magnetic field energy density, $n_e \sim B^2$. With this condition the change in electron density is dependent on magnetic field strength, or vice versa. Using $\Delta$|RM|= $387$~rad~m$^{-2}$, $n_e=0.014$~cm$^{-3}$, $B=0.07$~mG, we obtained the $\Delta B=0.2$~mG, and $\Delta n_e=0.028$~cm$^{-3}$.

For both of the above estimates of the electron density and magnetic field, the fluctuation will be much larger if the path length $l$ is the size of $\sim$17~pc, or of the same order of magnitude. Given the uncertainties of different estimates, long term component monitoring would be crucial to definitely rule out a foreground cloud as Faraday screen.

Another alternative explanation for |RM| and sign variability is the change in viewing angle, either intrinsically or caused by acceleration \citep{osullivan2009}. Regarding the source kinematics, models such as linear, helical, or precessing trajectories have been explored \citep{liu2010,algaba2019}. The change in viewing angle is naturally expected if the jet is following a helical trajectory. Jet encountering ambient medium and being deflected will also change its propagating angle. The observed jet region is located right at the apparent bending position, where the jet downstream turns northwest direction. Here we cannot explore the kinematics of the jet region in more detail, which is only partially resolved with the current resolution. \citet{broderick2009} explored another alternative, which is the relativistic helical bulk motion of the Faraday rotating jet sheath. They showed that the |RM| and sign can change with viewing angle concerning the Faraday rotating medium, with an asymmetric transverse RM profile. Here we do not have sufficient resolution to obtain a reliable transverse RM profile. The jet RM in the years 2006, 2007, and 2008 do not show significant spatial structure, this could be intrinsic, or the low resolution smoothing out the structure. Better resolution, and sensitivity will greatly help us to uncover these possibilities.

\section{Conclusions}

In this investigation, we focus on the RM variability from 2004 to 2008. For this purpose, we used multifrequency polarization observation from 4~GHz to 43~GHz with the VLBA. We consistently detected RM in the core region in five years. For the jet region we only detected RM located at the jet feature about 3.5~mas from core, during 2006, 2007, and 2008. We also observe that the RM changes in both magnitude and sign, with no apparent transverse structure, except in the jet region of 2008. 

At the core region, the 43-22~GHz RM magnitude is consistent within at least three years of observation, but from 2007 onward the RM sign changes from negative to positive. For RM around 15~GHz, emerging component C10 contributed briefly to the linear $\lambda^2$ law in year 2006. The C10 flux density then decayed to below the core flux density in 2007 onward, hence deviation from the linear $\lambda^2$ law, and sign change are observed. The 2007 power index $a$ (for $a$ in RM$\propto \nu^a$) was found to be around 1.48$\pm$1.40, is consistent with $a\sim2$, suggesting equipartition condition in jet sheath. For the jet region, high fractional polarization and linear $\lambda^{2}$ law of EVPA were observed, hence the situation of external Faraday screen is more likely. Jet-medium interaction and foreground cloud are less likely to be the Faraday screen, based on the large fluctuation $n_e$ and magnetic field needed to reproduce the observed RM changes. Alternative explanations based on jet kinematics are discussed. Higher resolution and better cadence are needed for a more complete investigation of this aspect.

\section*{Acknowledgements}

The authors are very grateful for the fruitful discussion with M. Johnston-Hollitt on complex CLEANing. K. C. Lai deeply appreciates the support of MyBrainSc Scholarship, under Ministry of Higher Education Malaysia. K. C. Lai also acknowledges the mutual support from all members from the Radio Cosmology Laboratory, Universiti Malaya. This research has made use of data from the MOJAVE database that is maintained by the MOJAVE team (Lister et al. 2018). We thank the anonymous referees for useful comments that helped improving the artile.

\section{Data availability}
The data underlying this article are available in NRAO Data Archive, at \url{https://data.nrao.edu/portal/}, under project codes BG152, BK107f, BL123B, BL137i, BG173b, and BA089. The VLBA 15~GHz data collected bt the MOJAVE programme are available at \url{https://www.cv.nrao.edu/MOJAVE/}. The VLBA 43~GHz data collected by the VLBA-BU-Blazar monitoring programme are available at \url{https://www.bu.edu/blazars/BEAM-ME.html}.



\bibliographystyle{mnras}
\bibliography{lai_RMvar1633_final} 




\appendix

\section{Data calibration of year 2005}

For the 2005 data, we combined two separate observations spaced apart by 3 days. Question may arise from the intraday variability of total intensity observed by \citet{volvach2009}. In the literature rapid variability of cm-wave VLBI polarization was largely observed for BL Lac object, from intraday variability \citep{gabuzda2000a, gabuzda2000b, gabuzda2000c, gabuzda1997}, to weeks \citep{charlot2006}. However no literature was found for polarization angle variablity for 1633+382. Long-term monitoring by the MOJAVE project \citep{lister2018b} had shown that throughout the year 2005 the 15~GHz EVPA had been consistent between $40^{\circ}\sim 47^{\circ}$. Hence we try to approximate the stability to 22 and 43~GHz in year 2005. 

\begin{figure}
	\includegraphics[width=\columnwidth]{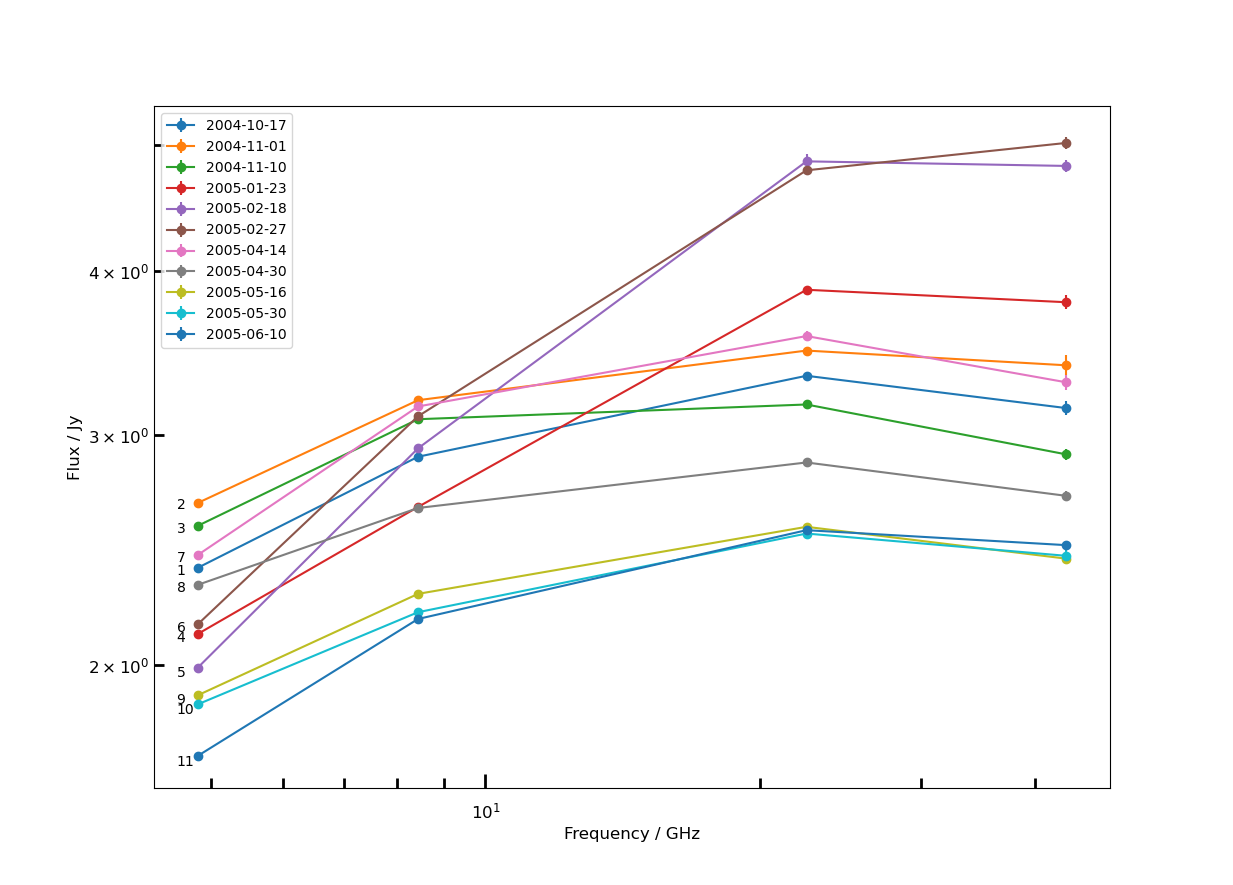}
    \caption{VLA integrated total intensity of 1749+096, for frequency 4.8, 804, 22.4, and 43.3~GHz. The label number 1 correspond to date 2004-10-17, and the subsequent labels follow the order of observation date. The spectral index rise started from 2005 Jan 23, reaching largest spectral index $\alpha\sim0.4$ in end of Febuary 2005, then flatten at April 2005.}
    \label{fig:1749spix}
\end{figure}

\begin{figure}
	\includegraphics[width=\columnwidth]{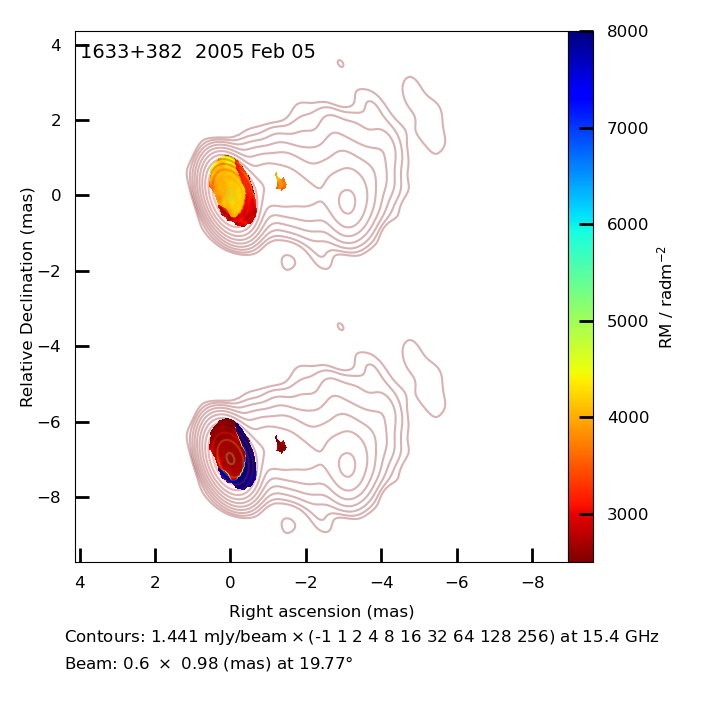}
    \caption{RM fitted without $90^{\circ}$ rotation. Upper plot is the RM map and lower plot is the RM error. The core rotation measure is roughly separated into region of $4500~rad m^{-2}$ and $3000~rad m^{-2}$, and the error at $2500~rad m^{-2}$ and $8000~rad m^{-2}$.}
    \label{fig:2005rmnorotateerr}
\end{figure}

The 15GHz data and CLEAN model were obtained from the MOJAVE website, under the project name BL123B. For 22~GHz and 43~GHz the calibration was done as in the standard AIPS calibration package for polarization, and EVPA correction using data from VLA POLCAL project. For 22 and 43~GHz a $90^{\circ}$ rotation was added to the final EVPA correction, due to the optically thin/thick transition \citep{gabuzda2001}. This is based on the suggested opacity change in EVPA calibrator 1749+096, as shown in Figure~\ref{fig:1749spix}. We assumed that the transition only affects the higher frequency, 22 and 43~GHz. This is further supported by the EVPA stability of 15~GHz throughout year 2005 as mentioned in the previous paragraph. Then for 1633+382, the RM fitted without $90^{\circ}$ rotation also shows gross discontinuity and large error, as in Figure~\ref{fig:2005rmnorotateerr}.

\section{Complex CLEANing}

The method of complex CLEANing was developed by \citet{pratley2016a}. Complex CLEAN is more superior than standard polarization CLEAN method in that it is rotationally invariant, detects more components at low signal-to-noise, and also fewer spurious components. We used the final calibrated data of 1633+382 from project BA089 and BL137I, convolved with their respective beam as in Table~\ref{tab:project_table} as our test data. The complex CLEANing was carried out in MIRIAD. Then both set of data was run through the same Rician debiasing process and then blanked. We compared the integrated EVPA, EVPA test point at core, and jet. The test point was chosen though eyeballing the polarization intensity peak in core and jet. Noted that the test value was taken from the same position, both in standard and complex CLEAN method. As shown in Table~\ref{tab:cclean_table}, the EVPA difference between two CLEANings is within the error. Although the difference in integrated EVPA is larger, only in 2 instances of project BA089 the difference is larger than $10^{\circ}$. We attributed the small differences to high SNR, and the compact nature of our source component. Hence we concluded the standard polarization CLEAN method would not introduced significant error in this study. 

\begin{figure*}
	\includegraphics{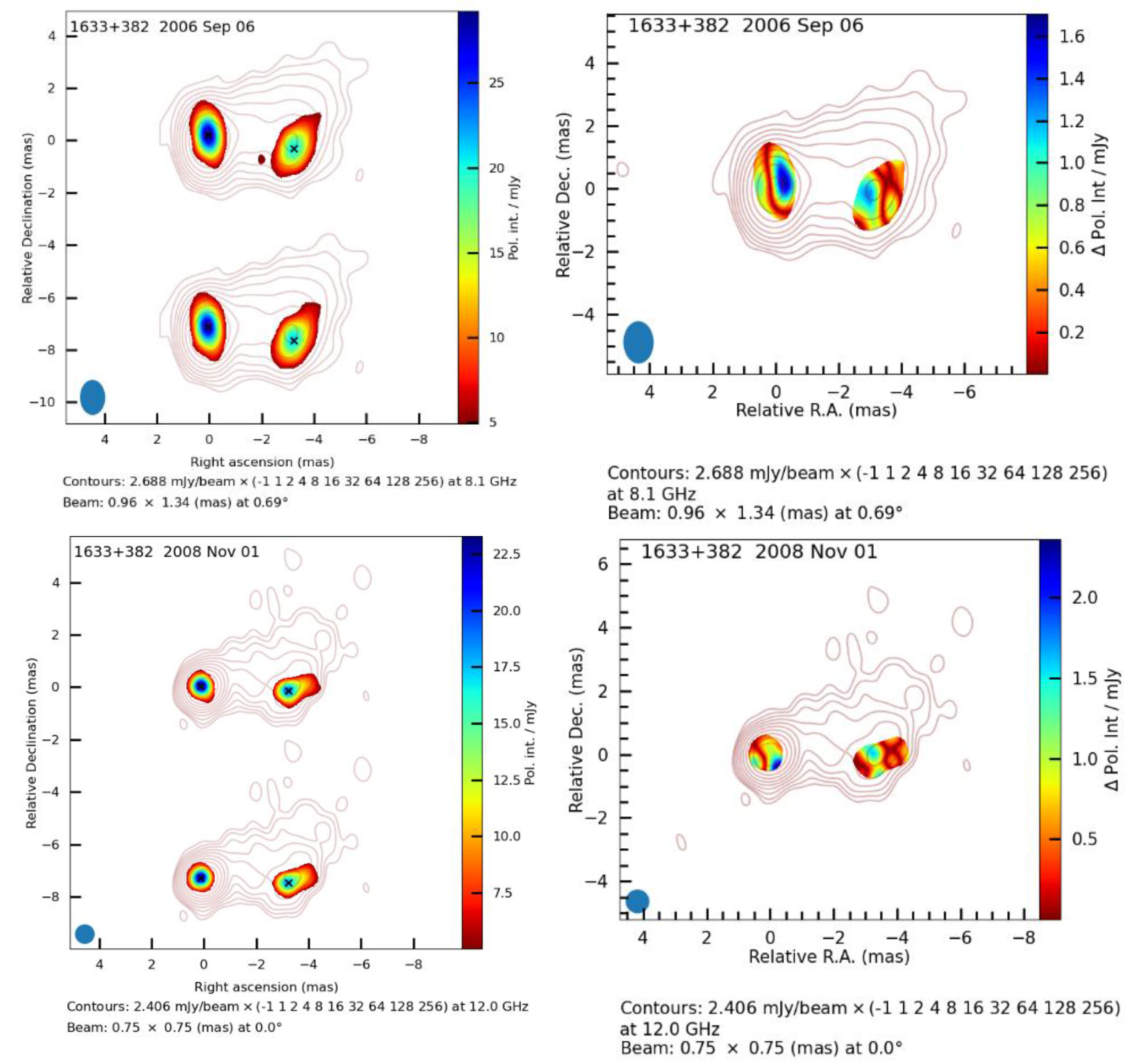}
	\caption{\textit{Left column:} BL137I 8.1~GHz (top) and BA089 12.0~GHz (bottom) was shown. In a single frame, the colour scale image on above is the polarized intensity image for complex CLEAN method, at below is the standard polarization CLEAN method. The same total intensity contour was plotted for both colour images to facilitate comparison. The black cross at the center of core and jet indicates the position where the test values were taken. Beam size was represented as the circle at bottom left corner. \textit{Right column:} The absolute value of the difference between complex CLEAN and standard polarization CLEAN is shown in colour scale.}
	\label{fig:cclean}
\end{figure*}

\begin{table*}
	\centering
	\caption{Result from standard polarization CLEANing and complex CLEANing}
	\label{tab:cclean_table}
	\begin{tabular}{cc|cccc|cccc|cccc} 
		\hline\hline
		Project & CLEAN & \multicolumn{4}{c}{$\chi_{integrated}$ ($^{\circ}$)} & \multicolumn{4}{c}{$\chi_{test,core}$ ($^{\circ}$)} & \multicolumn{4}{c}{$\chi_{test,jet}$ ($^{\circ}$)} \\ 
		 & & $\chi_{12}$ & $\chi_{15}$ & $\chi_{22}$ & $\chi_{24}$ & $\chi_{12}$ & $\chi_{15}$ & $\chi_{22}$ & $\chi_{24}$ & \\
		\hline
		BL137I & Standard & $39\pm3$ & $41\pm3$ & $58\pm6$ & $75\pm7$ & $31\pm1$ & $34\pm2$ & $52\pm2$ & $68\pm2$ & $51\pm2$ & $52\pm2$ & $67\pm2$ & $78\pm2$ \\
		 & Complex & $43\pm3$ & $44\pm3$ & $62\pm6$ & $76\pm7$ & $32\pm2$ & $34\pm2$ & $53\pm2$ & $68\pm2$ & $53\pm2$ & $52\pm2$ & $67\pm2$ & $75\pm3$ \\
		\hline
		BA089 & Standard & $61\pm4$ & $52\pm5$ & $57\pm8$ & $49\pm4$ & $65 \pm 3$ & $61 \pm 3$ & - & $54 \pm 5$ & $63 \pm 2$ & $63 \pm 2$ & $62 \pm 4$ & $60 \pm 3$ \\
		 & Complex & $65\pm4$ & $54\pm5$ & $74\pm8$ & $61\pm4$ & $67\pm3$ & $61\pm3$ & - & $68\pm5$ & $66\pm2$ & $63\pm2$ & $61\pm4$ & $63\pm3$ \\
		\hline
	\end{tabular}
\end{table*}

\section{EVPA and polarized intensity maps}
\begin{figure*}
	\includegraphics[scale=0.8]{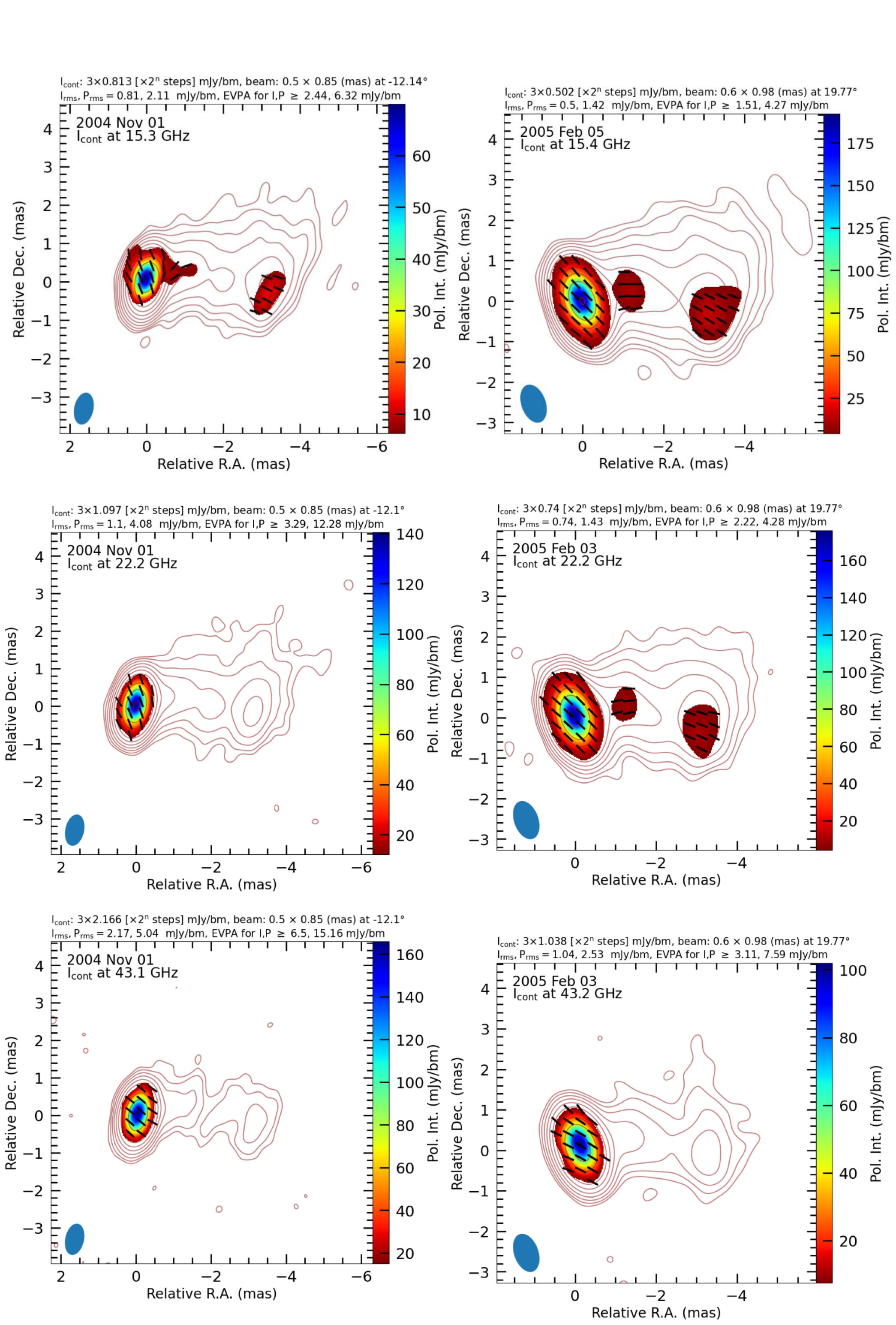}
	\caption{Tick mark of uniform length representing the observed EVPA, overlayed on polarized intensity colour map, at the indicated frequency. The polarized intensity was blanked on $3\sigma$ level. The contours show the Stokes I at $3\sigma \times 2^n$ steps.}
\end{figure*}

\begin{figure*}
	\includegraphics[scale=0.8]{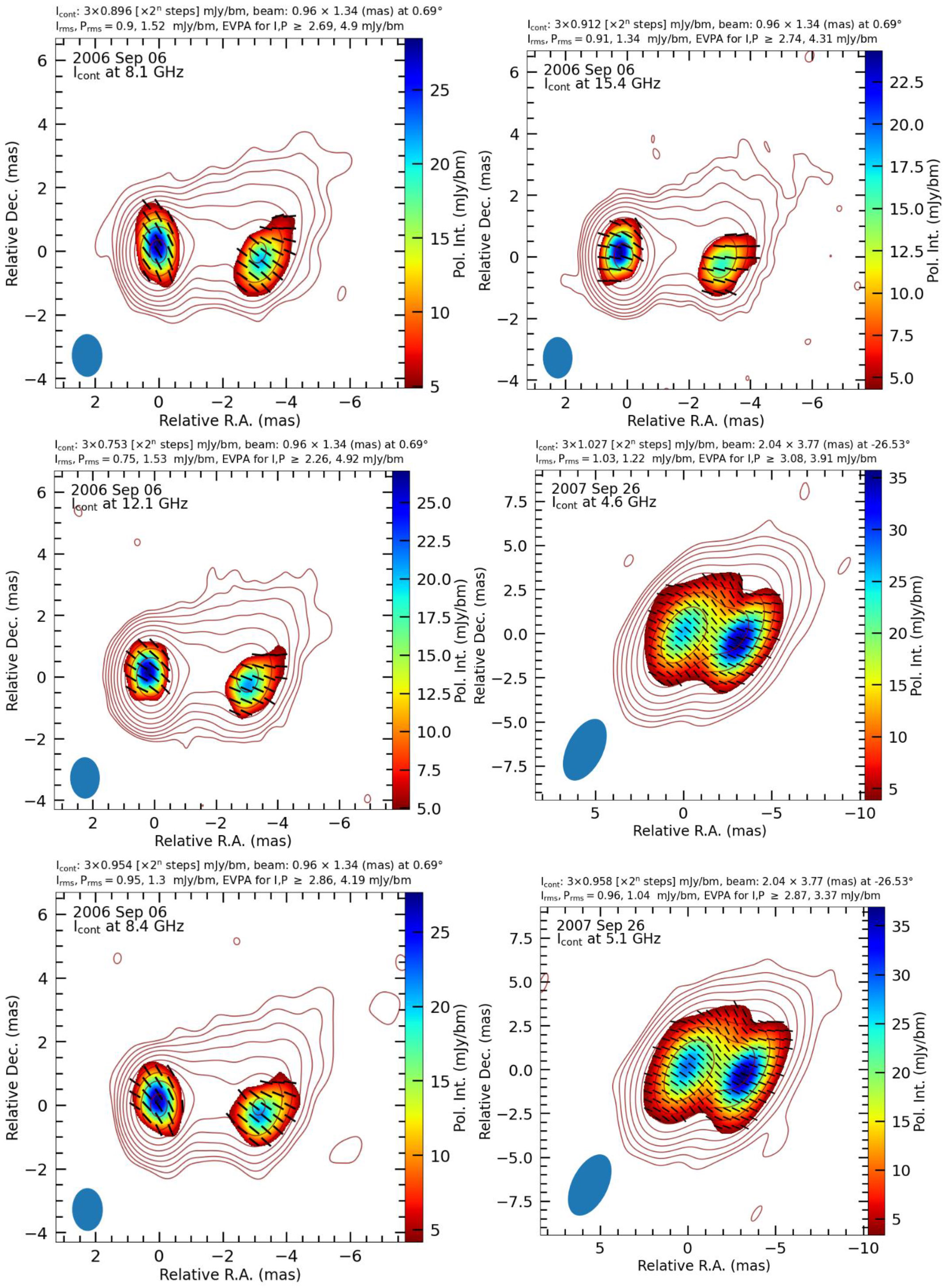}
	\caption{Cont.}
\end{figure*}

\begin{figure*}
	\includegraphics[scale=0.8]{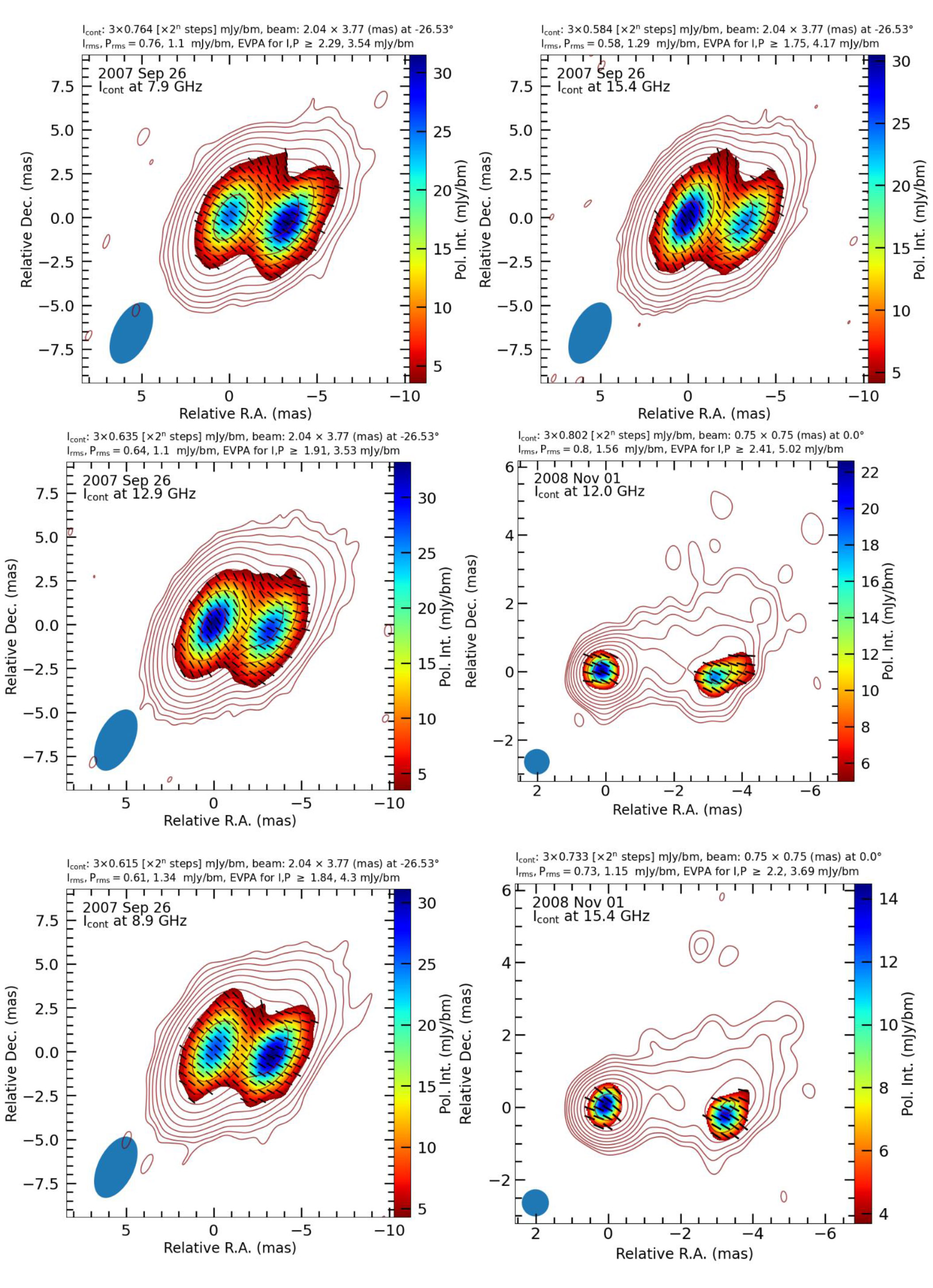}
	\caption{Cont.}
\end{figure*}

\begin{figure*}
	\includegraphics[scale=0.5]{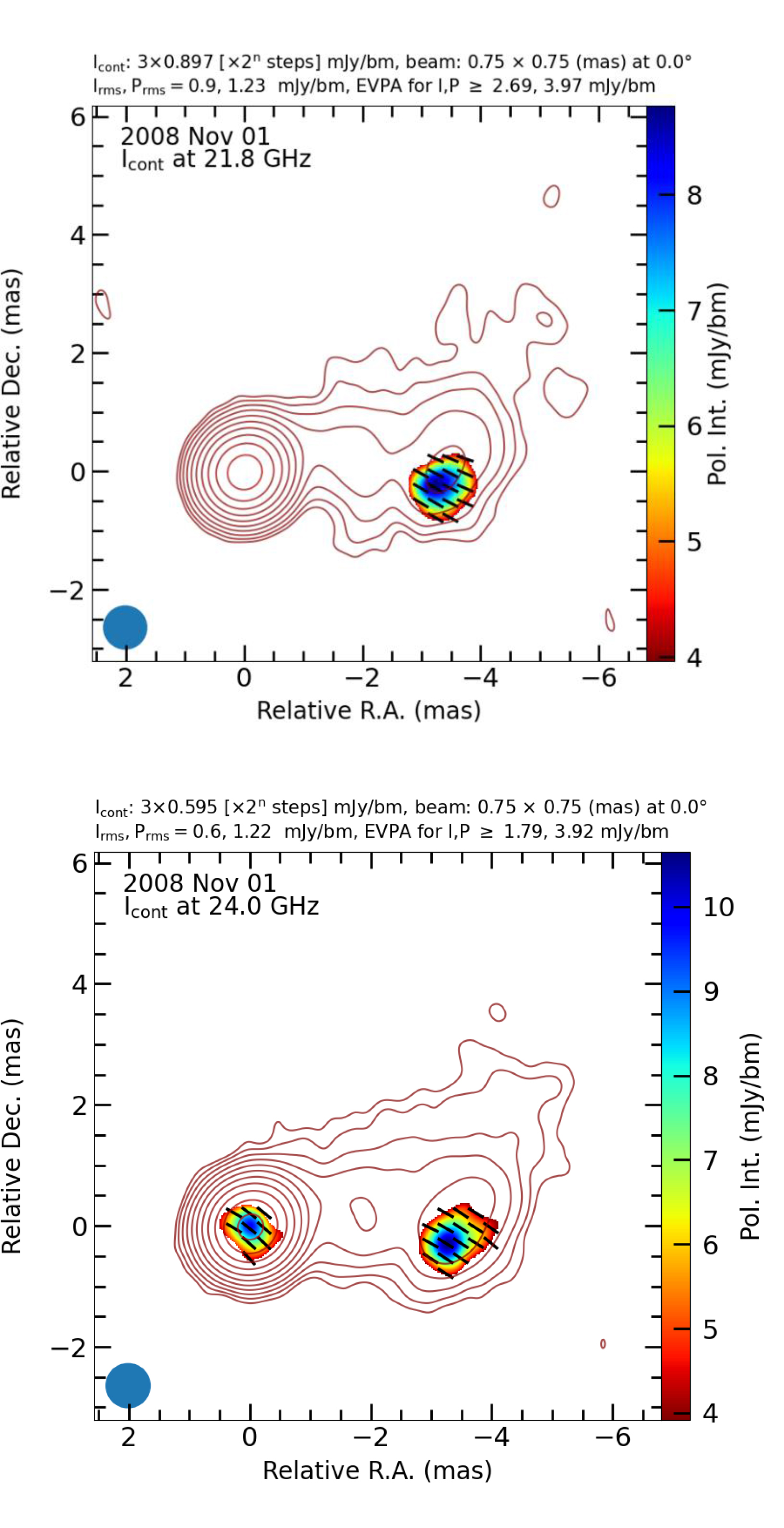}
	\caption{Cont.}
\end{figure*}

\section{Fractional polarization maps}
\begin{figure*}
	\includegraphics[scale=0.8]{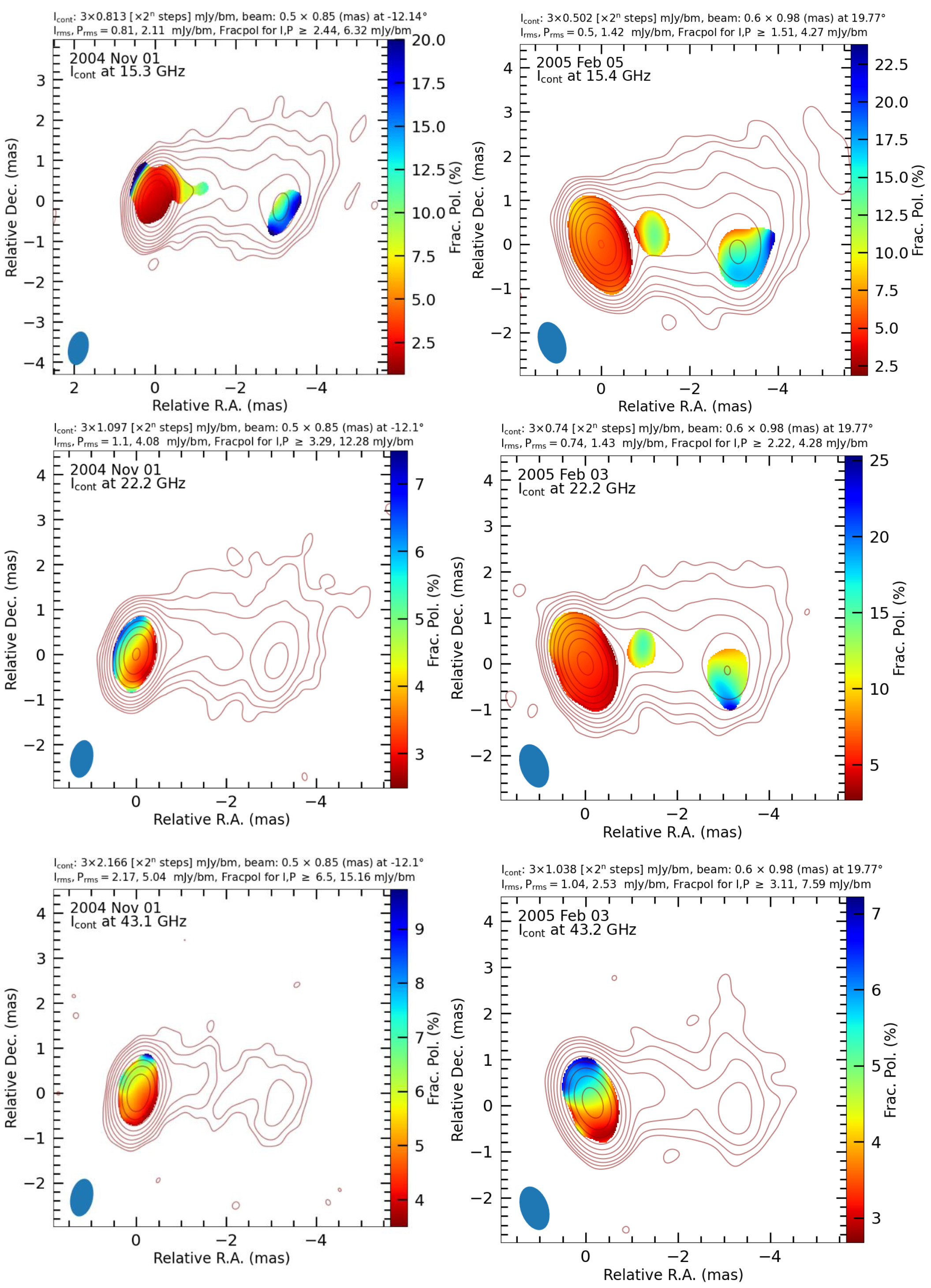}
	\caption{Fractional polarization maps, and the contours represent Stokes I. Note that in some maps the fractional polarization gradient at the core region is not seen because of its low value compared to the jet region.}
\end{figure*}

\begin{figure*}
	\includegraphics[scale=0.8]{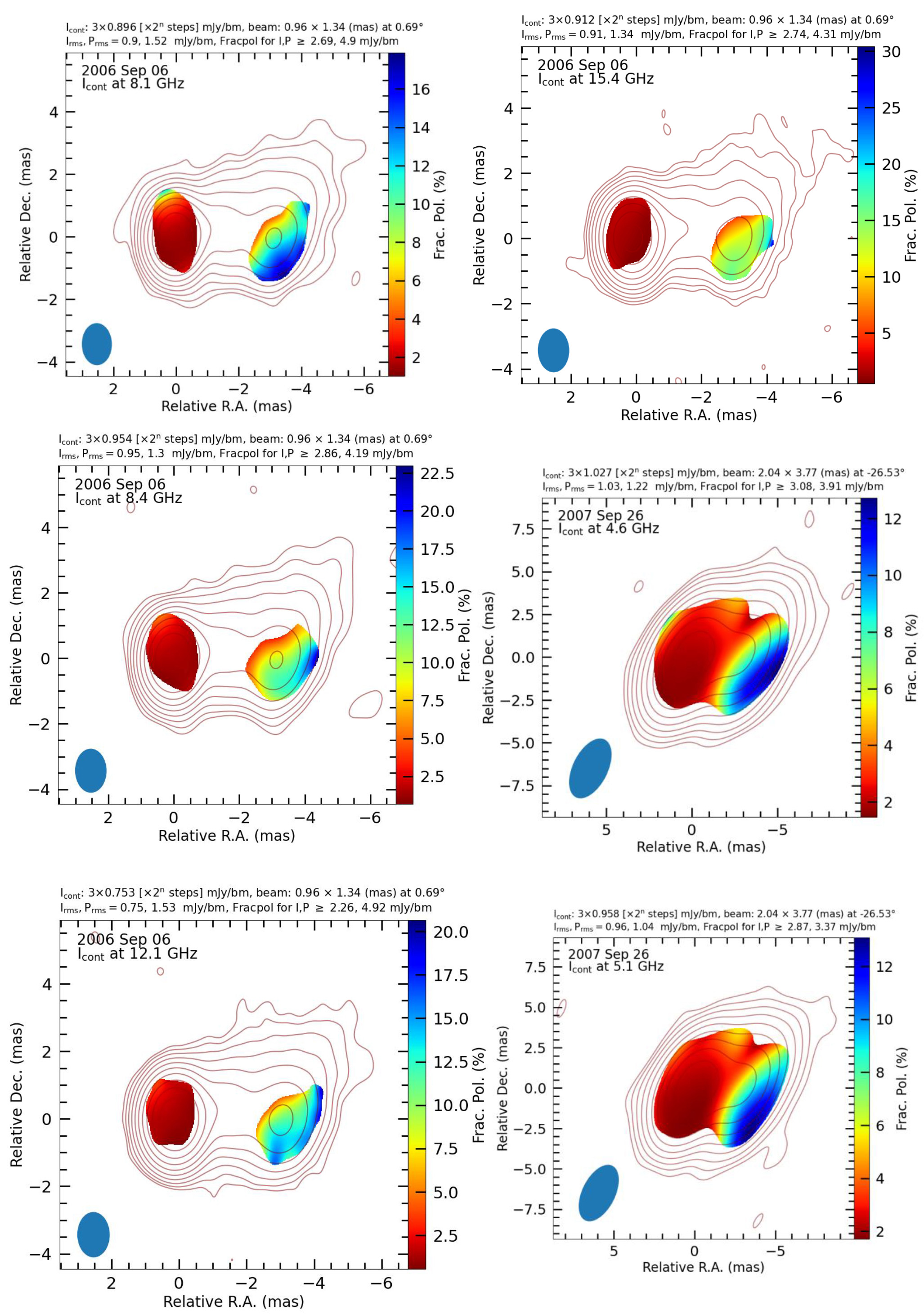}
	\caption{Cont.}
\end{figure*}

\begin{figure*}
	\includegraphics[scale=0.8]{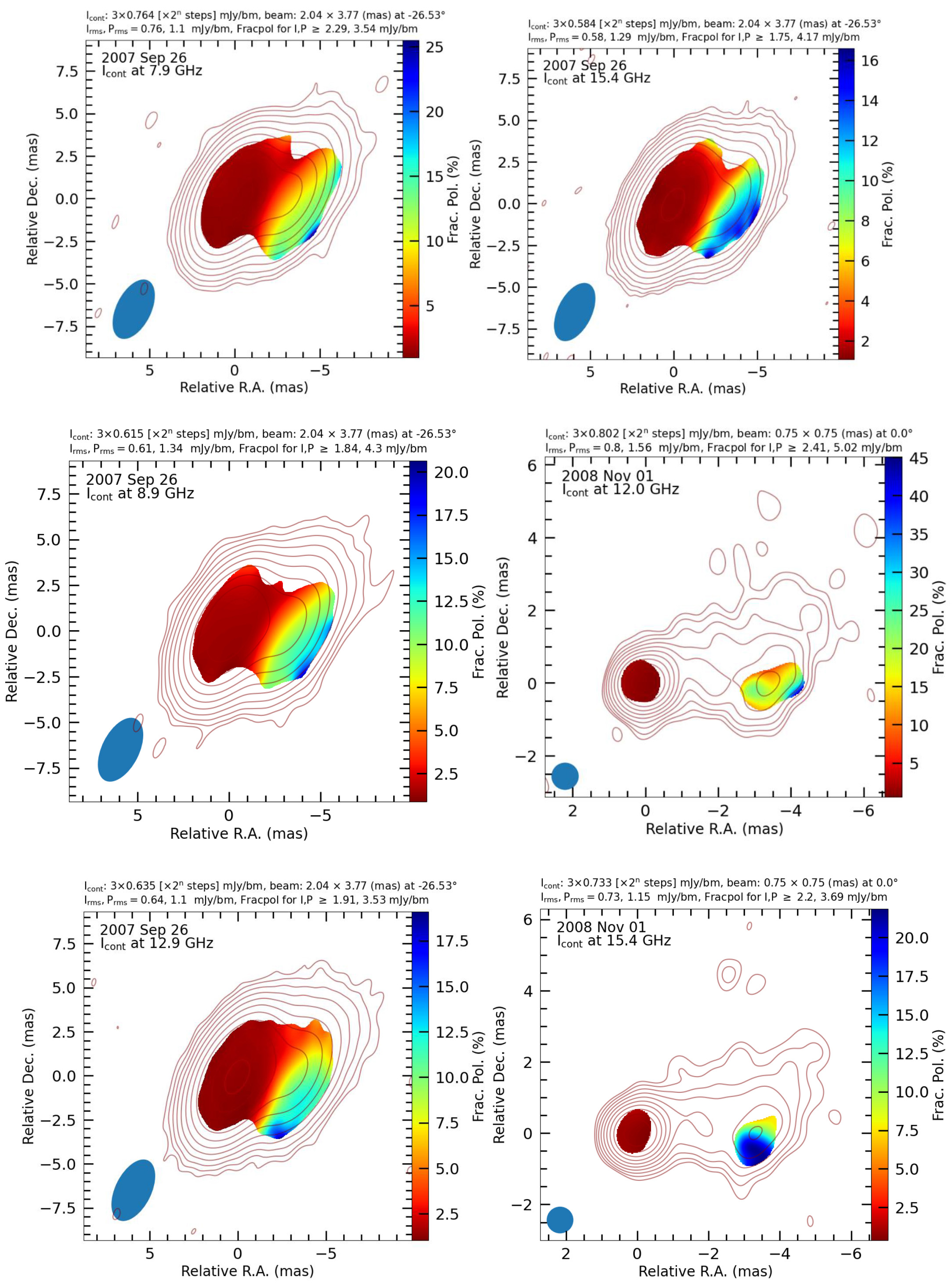}
	\caption{Cont.}
\end{figure*}

\begin{figure*}
	\includegraphics[scale=0.5]{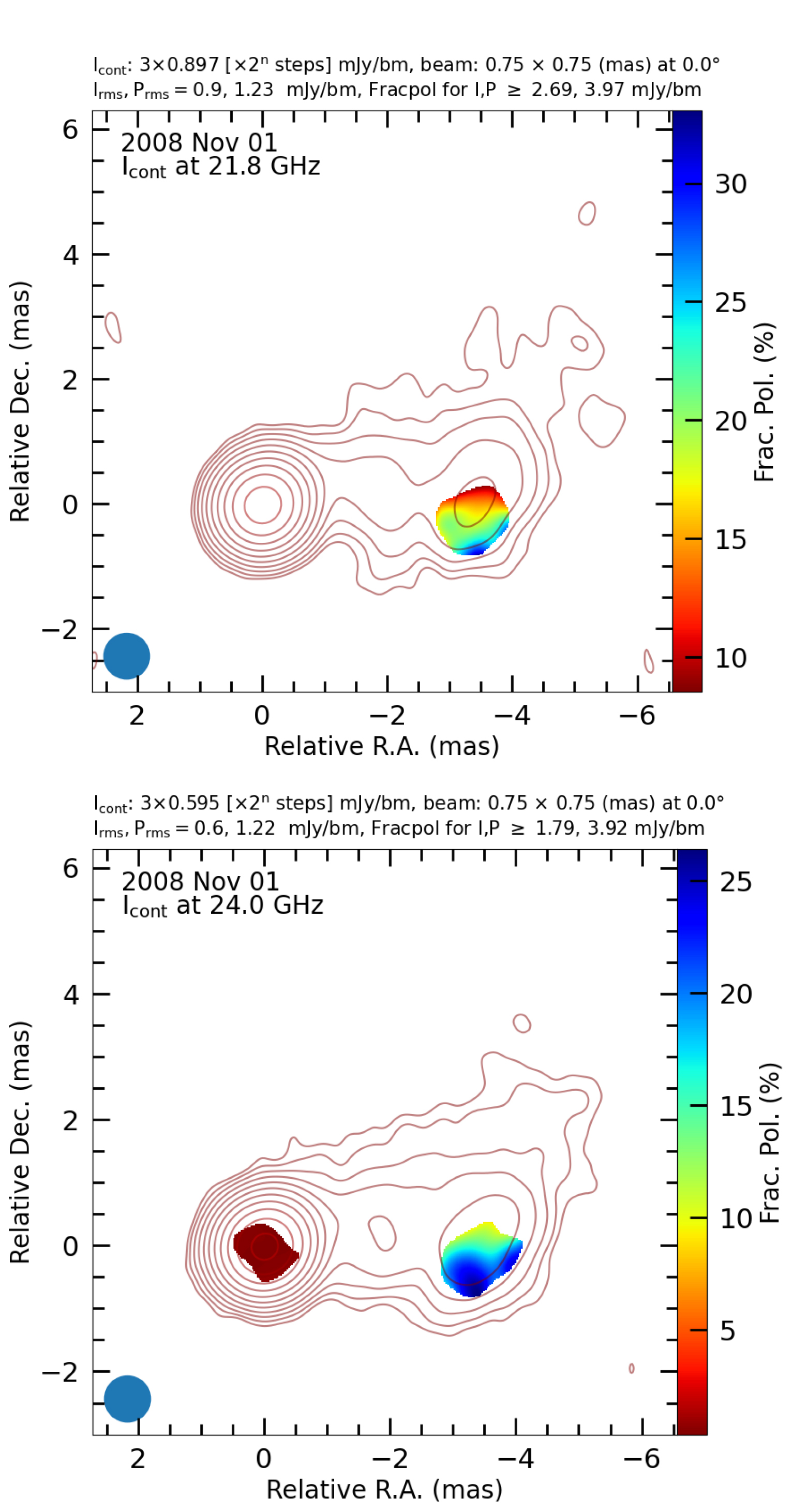}
	\caption{Cont.}
\end{figure*}


\bsp	
\label{lastpage}
\end{document}